\documentclass[apj]{emulateapj}

\lefthead{MODICA ET AL.} 
\righthead{GOALS OBSERVATIONS OF IC 883}

\begin{document}

\title{Multi-wavelength GOALS Observations of Star Formation and AGN
Activity in the Luminous Infrared Galaxy IC 883}

\author{F. Modica\altaffilmark{1},
T. Vavilkin\altaffilmark{1,2},
A. S. Evans\altaffilmark{1,2,3,4,5},
D.-C. Kim\altaffilmark{2,3},
J. M. Mazzarella\altaffilmark{6},
K. Iwasawa\altaffilmark{7},
A. Petric\altaffilmark{8},
J. H. Howell\altaffilmark{8},
J. A. Surace\altaffilmark{8},
L. Armus\altaffilmark{8},
H. W. W. Spoon\altaffilmark{9},
D. B. Sanders\altaffilmark{10},
A. Wong\altaffilmark{3},
\& J. E. Barnes\altaffilmark{10}
}

\altaffiltext{1}{Department of Physics \& Astronomy, Stony Brook
University, Stony Brook, NY, 11794-3800: aevans@virginia.edu;
tvavilk@vulcan.ess.sunysb.edu}

\altaffiltext{2}{National Radio Astronomy Observatory, 520 Edgemont Road,
Charlottesville, VA 22903}

\altaffiltext{3}{Department of Astronomy, University of Virginia, 530
McCormick Road, Charlottesville, VA 22904: dk3wc@virginia.edu;
aww2b@virginia.edu}

\altaffiltext{4}{Visiting Astronomer at the Infrared Processing and
Analysis Center, California Institute of Technology, MS 100-22, Pasadena,
CA 91125}

\altaffiltext{5}{Visiting Astronomer at the Institute of Astronomy and
Astrophysics, Academia Sinica, P.O. Box 23-141, Taipei 106, Taiwan, R.O.C.}

\altaffiltext{6}{Infrared Processing and Analysis Center, California
Institute of Technology, MS 100-22, Pasadena, CA 91125:
mazz@ipac.caltech.edu}

\altaffiltext{7}{ICREA and Institut de Ci\`encies del Cosmos (ICC), Universitat de 
Barcelona (IEEC-UB), Mart\'i i Franqu\`es, 1, 08028 Barcelona, Spain: 
kazushi.iwasawa@icc.ub.edu}

\altaffiltext{8}{Spitzer Science Center, Pasadena, CA 91125:
jhhowell@ipac.caltech.edu; jason@ipac.caltech.edu; lee@ipac.caltech.edu}

\altaffiltext{9}{Department of Astronomy, Cornell University, Ithaca, NY
14853: spoon@isc.astro.cornell.edu}

\altaffiltext{10}{Institute of Astronomy, University of Hawaii, 2680
Woodlawn Dr., Honolulu, HI 96822: sanders@ifa.hawaii.edu;
barnes@ifa.hawaii.edu}

\begin{abstract}

New optical {\it Hubble Space Telescope (HST)}, {\it Spitzer Space
Telescope}, {\it GALEX}, and {\it Chandra} observations of the
single-nucleus, luminous infrared galaxy (LIRG) merger IC 883 are
presented. The galaxy is a member of the Great Observatories All-sky LIRG
Survey (GOALS), and is of particular interest for a detailed examination of
a luminous late-stage merger due to the richness of the optically-visible
star clusters and the extended nature of the nuclear X-ray, mid-IR, CO and radio
emission.  In the {\it HST} ACS images, the galaxy is shown to contain 156
optically visible star clusters distributed throughout the nuclear
regions and tidal tails of the merger, with a majority of visible
clusters residing in an arc $\sim$ 3--7 kpc from the position
of the mid-infrared core of the galaxy. The luminosity functions of the
clusters have an $\alpha_{\rm F435W} \sim -2.17\pm0.22$ and
$\alpha_{\rm F814W} \sim -2.01\pm0.21$, compared with V-band derived values
measured for the well-studied LIRG NGC 34 and the Antennae Galaxy of
$\alpha \sim -1.7\pm0.1$ and $-2.13\pm0.07$, respectively. Further, the
colors and absolute magnitudes of the majority of the clusters
are consistent with instantaneous burst population synthesis model ages in
the range of a few$\times10^7 - 10^8$ yrs (for $10^5$ M$_\odot$ clusters),
but may be as low as few$\times10^6$ yrs with extinction factored in.
The X-ray and mid-IR spectroscopy are indicative of predominantly
starburst-produced nuclear emission, and the star formation rate, estimated
based on the assumption that the radio and far-infrared luminosities are
tracing the starburst population, is $\sim$ 80 M$_\odot$ yr$^{-1}$. The
kinematics of the CO emission and the morphology of both the CO and radio
emission are consistent with the nuclear starburst being situated in a
highly inclined disk 2 kpc in diameter with an infrared surface brightness
$\mu _{\rm IR} \sim 2\times10^{11}$ L$_\odot$ kpc$^{-2}$, a factor of 10
less than that of the Orion star-forming region.  Finally, the detection of
the [Ne V] 14.32$\mu$m emission line is evidence that an AGN is present.
The faintness of the line (i.e., [Ne V] / [Ne II] 12.8$\mu$m $\sim 0.01$)
and the small equivalent width of the 6.2$\mu$m PAH feature ($= 0.39\mu$m)
are both indicative of a relatively weak AGN.

\end{abstract}


\section{Introduction}

The $z = 0.0233$ galaxy IC 883 (= UGC 8387, Arp 193, B2 1318+343; $L_{\rm
IR}$ [8--1000$\mu$m] $\sim 5.4\times10^{11}$ L$_\odot$) is a member of the
luminous infrared galaxies class (LIRGs:  $L_{\rm IR} \geq 10^{11}$
L$_\odot$: e.g., Sanders \& Mirabel 1996).  As is the case with some of the
local LIRG population, the high $L_{\rm IR}$ of IC 883 is the result of a
recent galaxy-galaxy merger event -- the galaxy is observed optically to be
a single-nucleus, advanced merger with two tidal tails. It is optically
classified as a LINER (Kim et al. 1995; Veilleux et al. 1995), with the
evidence thus far supporting star formation as its primary energy source.
Millimeter (CO) observations of the galaxy show it to have $\sim
4\times10^{9}$ M$_\odot$ of molecular gas (adopting $M[{\rm H_2}] /
L^{\prime}_{\rm CO} = 0.8$ M$_\odot$ [K km s$^{-1}$ Mpc$^2$]$^{-1}$: Downes
\& Solomon 1998) within the  inner several kpc; the morphology and velocity
profile of the gas is consistent with it being situated in a highly
inclined, rotating disk (Downes \& Solomon 1998; Bryant \& Scoville 1999).
Further, the {\it HST} NICMOS (Scoville et al. 2000) and Keck mid-IR
(Soifer et al. 2001) imaging data also show evidence of a bright edge-on
disk $\sim 4\arcsec$ (2 kpc) in length in the nuclear region of the merger.

In this paper, new {\it Hubble Space Telescope} ({\it HST}), {\it Spitzer
Space Telescope}, {\it Chandra} X-ray observatory, {\it GALEX}, and Owens
Valley Millimeter Array (OVRO) observations, and archival VLA and {\it HST}
NICMOS data, of IC 883 are presented. The galaxy is a member of the Great
Observatories All-sky LIRG Survey (GOALS: Armus et al. 2009), which is a
multi-wavelength survey of a complete sample of LIRGs in the {\it IRAS}
Revised Bright Galaxy Sample (RBGS: Sanders et al. 2003); the present paper
is the third in a series focussing on a single merger system. In the
particular case of IC 883, the presence in a late-stage merger of $ > 100$
optically-visible star clusters, combined with the spatial
extent of the nuclear X-ray, mid-IR, CO and radio emission, makes it
ideally suited for examining the properties of distributed
optically-visible star formation, and for constraining the nuclear activity
in IC 883.

This paper is divided into six sections. In \S 2, the observations and data
reduction are summarized. Section 3 contains a general overview of the
results, followed by a discussion of the cluster identification and
photometric methods in \S 4 (for a more detailed description of cluster
identification and photometry, see Vavilkin et al. 2011, in preparation).
Section 5 covers two major topics. First, the optical properties
(distribution, luminosity function and ages) of the {\it HST}-visible star
clusters are derived. Second, the discussion is expanded to include the
broader GOALS dataset, thus placing the optical observations, and the
observed activity in IC 883, into the proper context. A summary of the
findings (\S 6) concludes the paper.

The galaxy IC 883 has a heliocentric radial velocity of $6985 (\pm 3)$ km
s$^{-1}$ (Rothberg \& Joseph 2006).  The velocity, corrected for
perturbations by the Virgo Cluster, the Great Attractor and the Shapley
Supercluster (Mould et al. 2000), is $7000 (\pm 15)$ km s$^{-1}$.
Correcting the corresponding $z$ for Hubble flow in a cosmology with $H_0 =
70$ km s$^{-1}$ Mpc$^{-1}$, $\Omega_{\rm m} = 0.28$, and $\Omega_\Lambda =
0.72$ (Hinshaw et al. 2009) results in a luminosity distance to IC 883 of
110 Mpc and a scale of 0.507 kpc/$\arcsec$ (see also Armus et al. 2009).

\section{Observations and Data Reduction}

{\it Hubble Space Telescope} observations of IC 883 with the Wide Field
Channel (WFC) of the Advanced Camera for Surveys (ACS) were obtained on
2006 January 11 as part of a much larger campaign (HST-GOALS: Evans et al.
2011, in preparation) to image a complete sample of 88 $L_{\rm IR} \geq
10^{11.4}$ L$_\odot$ {\it IRAS} galaxies in the Revised Bright Galaxy
Sample (RBGS: Sanders et al.  2003).  The galaxy was imaged with the  F435W
and F814W filters in a single orbit, with integration times of 1290 and 740
seconds, respectively.  The data were reduced in the manner described in
Evans et al. (2011).

The {\it Spitzer} IRAC and MIPS observations were obtained on 2005 June 11
and June 20, respectively, as part of the {\it Spitzer} component of GOALS
(Mazzarella et al. 2011, in preparation).  Integration times correspond to
150 seconds for the IRAC 3.6$\mu$m, 4.5$\mu$m, 5.8$\mu$m, and 8$\mu$m
channels and 48, 38, and 25 seconds for the MIPS 24$\mu$m, 70$\mu$m, and
160$\mu$m channels, respectively.  The {\it Spitzer} data were reduced in
the manner described in detail in Mazzarella et al. (2011).

The high resolution {\it Spitzer} IRS data on IC 883 was taken as part of a
Cycle 2 {\it Spitzer} observing campaign (PID 20549 AOR key 1483926). The
source was observed with the high resolution modules (120 seconds $\times$
1 cycles in SH and 60 seconds $\times$ 1 cycles in LH). No offsets were
taken in this program. The low resolution SL data (120 seconds $\times$ 2
cycles) were taken as part of {\it Spitzer} observing campaign (PID 3237
AOR key 10509824).  The starting point for the data reduction was the Basic
Calibrated Data (BCDs) processed by the SSC pipeline (ver. S15.3.0).  The
SL and SH modules were first corrected for rogue pixels using an automatic
search-and-interpolate algorithm (which was supplemented by visual
examination).  In order to estimate the background in SL module, the two
nods were subtracted.  To extract 1-D spectra, the Spectroscopic Modeling
Analysis and Reduction Tool (SMART; Higdon et al. 2004) was used.  The
extraction apertures were tapered with wavelength to match the point-spread
function for SL data and encompassed the entire slit for SH data.  For SL,
the two one-dimensional nod spectra for the orders SL1 and SL2 were
combined separately and then the orders were stitched by trimming a few
pixels from one or the other order.  For SH, the two nods were combined and
then stitched together by applying multiplicative offsets for each order
that were linear in flux density vs. wavelength (effectively removing a
tilt artifact from certain orders where necessary).  The zodiacal light in
SH spectra was subtracted using a blackbody fit to the Spitzer Planning
Observations Tool (SPOT) zodiacal estimates at 10$\mu$m and 20$\mu$m.

The {\it GALEX} far-UV ($\lambda_{\rm eff} = 0.1528\mu$m) and near-UV
($\lambda_{\rm eff} = 0.2271\mu$m) imaging observations were done as part
of the {\it GALEX} All-Sky Survey (AIS). The data were obtained on 2004
April 04, with an integration time of 119 seconds per band and were reduced
using the standard {\it GALEX} pipeline.

The {\it Chandra} X-ray Observatory observations were obtained on 2007
February 19. The total integration time of the observations was 14.07 ks,
and the data were reduced using the standard pipeline.

The CO($1\to0$) observations of IC 883 were obtained during three observing
periods from 1999 December--2000 January with the Owens Valley Millimeter
Array (OVRO). The array consists of six 10.4m telescopes, and the longest
observed baseline was 242m. Each baseline was configured with 120$\times$4
MHz digital correlators. During the observations, the nearby quasar
1308+326 was observed every 25 minutes to monitor phase and gain
variations, and 3C 273 was observed to determine the passband structure.
Finally, observations of Uranus were made for absolute flux calibration.
The OVRO data were reduced and calibrated using the standard Owens Valley
Millimeter Array program (MMA: Scoville et al. 1993). The data were then
exported to the mapping program DIFMAP (Shepherd, Pearson, \& Taylor 1995)
for extraction of the intensity map and spectra.

\begin{figure*}[htp]
\centerline{\includegraphics[angle=-90,width=\textwidth]{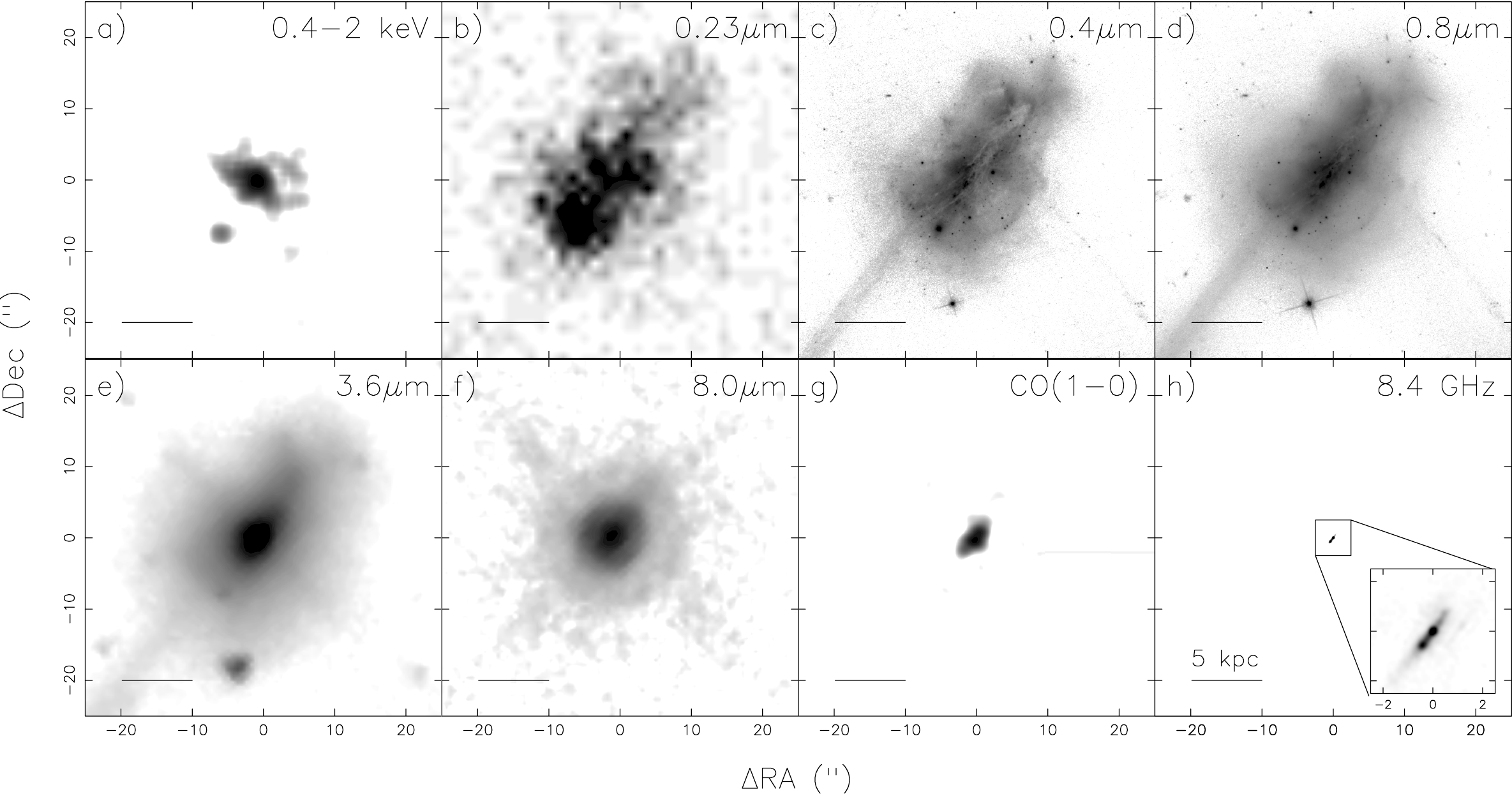}}
\figcaption{Multi-wavelength view of the nuclear region and main body of IC
883 displayed in order of increasing wavelength. The images have been
spatially registered such that $\Delta$RA = 0 and $\Delta$Dec = 0
correspond to RA = 13:20:35.4 Dec = +34:08:22.6 (J2000). The images shown
are as follows: (a) Chandra, (b) GALEX, (c,d) HST, (e,f) Spitzer, (g) OVRO,
(h) VLA.  With the exception of the {\it GALEX} and VLA images, all of the
images are shown with a logarithmic stretch. The integrated intensity CO
map is shown with natural weighting with a beam FWHM of
$1.92\arcsec\times1.35\arcsec$ and a position angle of 79.7$^{\rm o}$.  In
all of the images, North is up, East is to the left.} 
\end{figure*}

The nuclear region of reduced images of IC 883 are shown in Figure 1 in
order of increasing wavelength. Note that, due to the similarity in the
morphology of IC 883 in the far- and near-UV, at 3.6 and 4.5 $\mu$m, and at
5.8 and 8.0 $\mu$m, only the near-UV, 3.6 $\mu$m and 8.0 $\mu$m images are
shown. Figure 1 also includes an 8.4 GHz radio image of IC 883 taken with the
VLA over the period 1990 February -- March (Project AH0396: Condon et
al. 1991). Figure 2 contains the whole galaxy at the near-UV wavelength, in
both {\it HST} filters, and at 3.6$\mu$m.

\begin{figure*}[htp]
\centerline{\includegraphics[angle=-90,width=5in]{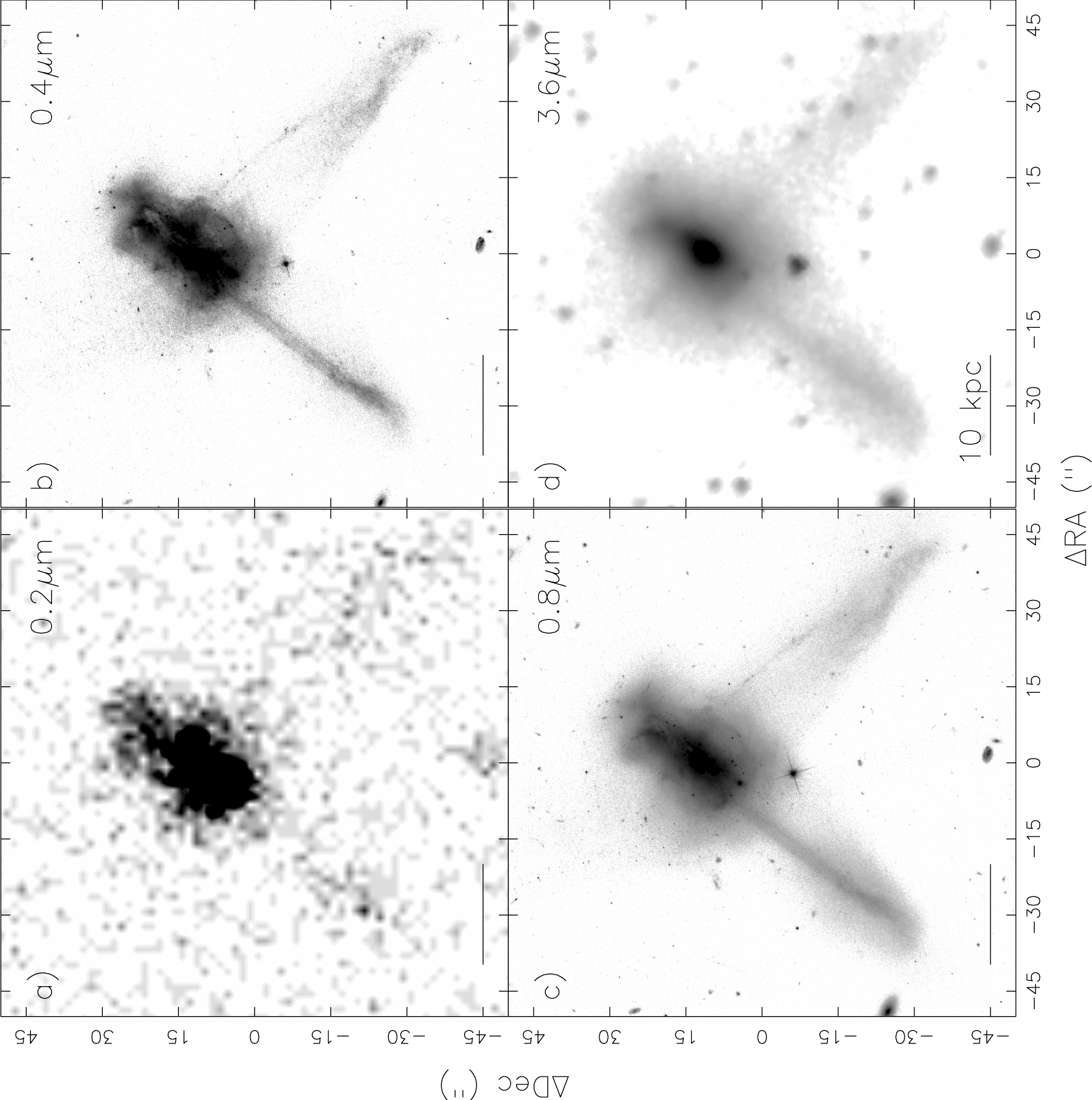}}
\figcaption{Multi-wavelength, wide-field view of IC 883 displayed in order
of increasing wavelength. The images shown are as follows: (a) GALEX, (b,c)
HST, (d) Spitzer.  With the exception of the {\it GALEX} image, all of the
images are shown with a logarithm stretch.  In all of the images, North is
up, East is to the left.} 
\end{figure*}

\section{Results}

The global appearance of the galaxy IC 883 is similar over the 0.1 --
3.6$\mu$m wavelength range (Figures 1 and 2), but its large-scale
morphology is easiest to assess in the {\it HST} ACS images due to their
depth and high resolution.  The galaxy consists of a diffuse, elongated
main body 17 kpc in length with two tidal tails extending in the
southeastern and southwestern directions (Figure 2b and 2c). The
southeastern tail is linear in appearance with a projected full extent of
20 kpc from the main body. In contrast, the southwestern tail extends from
the main body out to a projected distance of 27 kpc, then back towards the
main body where it eventually becomes difficult to distinguish against the
high surface brightness haze. In addition to the large scale structure
visible in the {\it HST} images, a small fraction of the more than 100
bright knots (see \S 4), which are presumably luminous star clusters
formed as a result of the merger, in IC 883 are shown in Figure 1c and 1d
to be distributed within the extended nuclear region and main body of the
galaxy.

Figure 1b shows the main body of IC 883 in the near-UV. The highest surface
brightness emission in the near-UV emanates from the southern half of its
main body, and is offset from the bright nuclear cores visible in the {\it
Spitzer} IRAC and MIPS images (Figure 1e and 1f).

At longer wavelengths, the inner few kpc of IC 883 make up the bulk of the
emission.  At 8$\mu$m, the galaxy consists of an unresolved nucleus which
is surrounded by a faint haze.  In the lower resolution MIPS images (not
shown), which cover the wavelength range 24--160$\mu$m, the galaxy appears
as a single, unresolved point source. The true compactness of the
IR-emitting region may be best represented by the CO(1$\to$0) and 8.4 GHz
radio emission images (Figures 1g and 1h: see discussion in \S 5.3). Both
the CO and radio emission have elongated morphologies, with full extents of
$\sim$ 2 kpc. Two high surface brightness knots, separated by 350 pc, are
observed in the radio image, one of which is spatially aligned with the
8$\mu$m core of IC 883.

A substantial amount of the literature exists on the CO properties of IC
883.  The measured CO flux density of the present dataset is $132\pm12$ Jy
km s$^{-1}$, with an additional calibration uncertainty which is likely to
be as high as 20\%.  By comparison, Downes \& Solomon (1998) and Bryant \&
Scoville (1999) report CO flux densities from their interferometric
observations of $161\pm32$ Jy km s$^{-1}$ and 202 Jy km s$^{-1}$,
respectively.  Morphologically, the appearance of CO synthesized intensity
map shown in Figure 1g is similar to those of Downes \& Solomon (1998) and
Bryant \& Scoville (1999). There are also several published single-dish
measurements of IC 883 (Mazzarella et al.1993; Downes \& Solomon 1998;
Evans et al. 2005); their reported CO flux densities are 220, 162, and 177
Jy km s$^{-1}$, respectively.

The X-ray emission is composed of extended emission centered on the 8$\mu$m
galaxy core and a weak point-like source, probably associated with a star
cluster to SE. The position of the primary extended emission centroid is
13h 20m 35.3s, +34d 08m 21.9s (J2000) and of the SE compact source is 13h
20m 35.7s, +34d 08m 14.3s (J2000).

The integrated counts of the primary X-ray component are 251.4$\pm$16.0
counts in the 0.4-7 keV band and 15.8$\pm$4.0 counts for the SE component.
The hardness ratio, defined as $HR=(H-S)/(H+S)$ where $H$ is $2-8$ keV
counts and $S$ is $0.5-2$ keV counts, of the two components are: $HR$(main)
= $-0.55\pm0.07$ and $HR$(SE) = $-0.64\pm0.30$.

The soft X-ray emission ($0.4-2$ keV: see Figure 1a) is extended in the
direction of NE-SW, nearly perpendicular to the major axis of the galaxy.
Very faint filamentary emission might extend up to 16$\arcsec$ (8 kpc) to
SW. The brightness distribution of the soft X-ray emission is rather flat:
when fitted with a King profile, it is characterised by a large core radius
of  10$\arcsec$ ($\sim 5$ kpc) and a steep index of 10 (indicating a rapid
decline beyond the core radius). The 2--7 keV emission is marginally resolved.

\begin{figure}
\centerline{\includegraphics[angle=-90,width=3in]{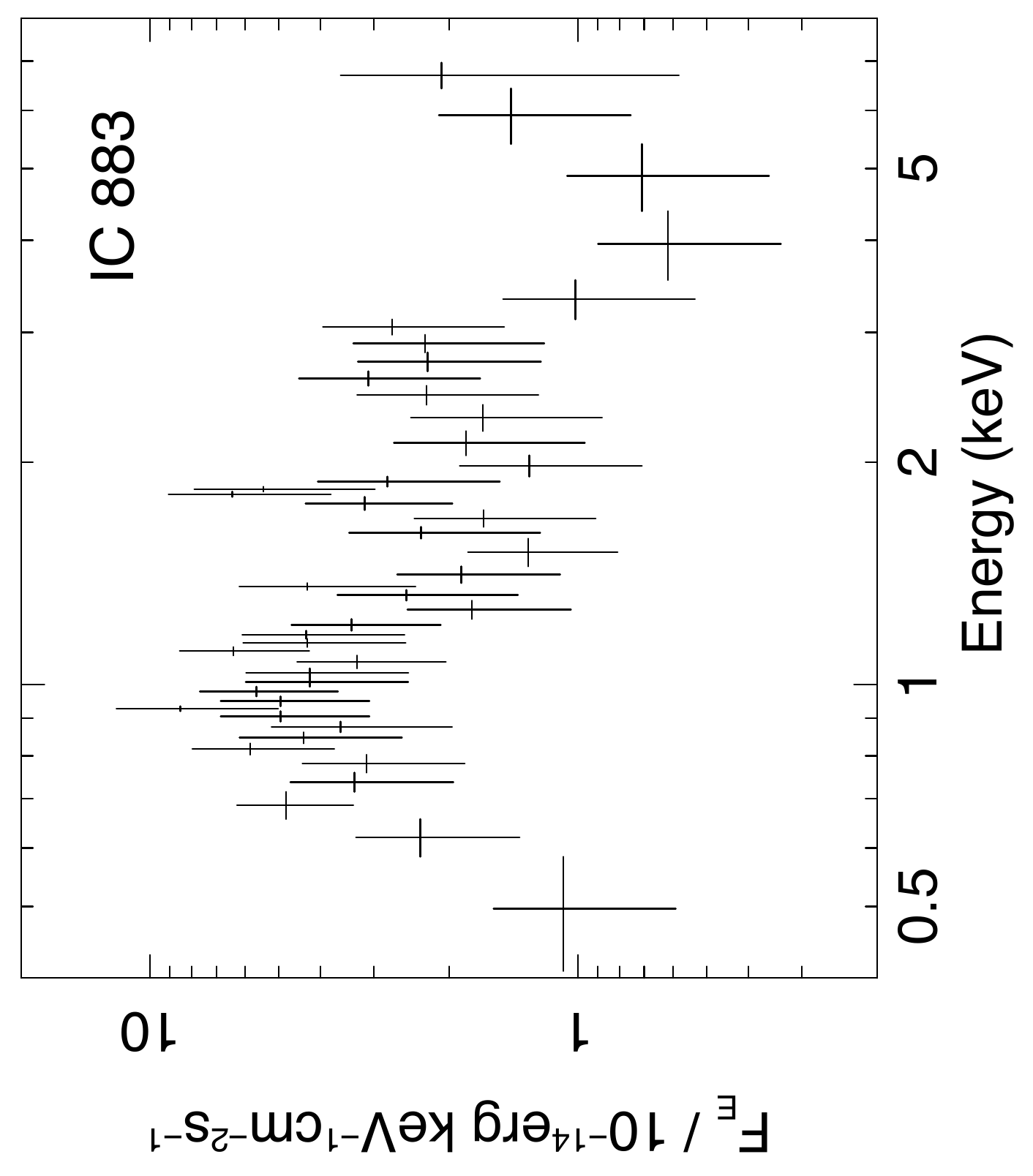}}
\figcaption{The response-corrected X-ray spectrum of IC 883 obtained from
the Chandra ACIS-S. The flux density is plotted in units of $10^{-14}$ erg
s$^{-1}$ keV$^{-1}$, and the energy is in units of keV in the observer
frame. The soft X-ray emission can be modeled as thermal emission with a
temperature of $kT \sim 0.7$ keV. A strong Si XIII line at 1.8 keV may
suggest $\alpha$-elements enriched gas composition. The hard X-ray tail
above 3 keV comes from the central compact source, and it spectral shape is
not well constrained due to small detected counts.} 
\end{figure}

The X-ray spectrum (Figure 3) was extracted only from the primary emission
region (the SE component is too faint to construct a spectrum for an
independent investigation). The spectrum consists of thermal emission in
the soft band.  There is a possible presence of a hard tail in the Chandra
spectrum above 4 keV. Given the limited detected counts, no good constraint
on the spectral shape can be obtained but its slope could be rather flat,
with a photon index $\Gamma\leq 1$.  The soft X-ray emission features can
be described by a thermal emission spectrum with a temperature of $kT
=1.0\pm 0.1$ keV with $\alpha $ element enhanced metallicity pattern. This
is consistent with an interstellar medium enriched by core-collapse
supernovae, as expected from a starburst. The Si XIII is very strong with
an equivalent width of 0.26 keV, which requires $\sim 3$ times more
metallicity than the other $\alpha $ elements. Since the Si XIII line is on
the higher energy side of the thermal spectrum, the enhancement could be
due to an extra contribution of AGN photoionized gas,
as suspected based on the similarly strong Si XIII line found in other GOALS 
galaxies hosting Compton thick AGN (Iwasawa et al 2011).  The soft band 
spectrum may likely be best explained by a
non-solar abundance ratio with enhanced $\alpha$-elements. A further
investigation is reported by Iwasawa et al. (2011). The observed (i.e.,
uncorrected for absorption) fluxes and luminosities of the nuclear and SE
source X-ray emission are summarized in Table 1.

\begin{deluxetable}{lrr}
\tablenum{1}
\tablewidth{0pt}
\tablecaption{Chandra X-ray Fluxes and Luminosities}
\tablehead{
\multicolumn{1}{c}{Band} &
\multicolumn{1}{c}{Flux} &
\multicolumn{1}{c}{Luminosity}
\nl
\multicolumn{1}{c}{(keV)} &
\multicolumn{1}{c}{(erg s$^{-1}$ cm$^{-2}$)} &
\multicolumn{1}{c}{(erg s$^{-1}$)}
}
\startdata

Nuclear & & \nl
0.5--2 & $4.2\pm0.4\times10^{-14}$ & $6.1\pm0.6\times10^{40}$ \nl
2--7    & $5.1\pm0.8\times10^{-14}$ & $7.4\pm1.2\times10^{40}$ \nl
2--10  & $6.5\pm1.2\times10^{-14}$ & $9.4\pm1.2\times10^{40}$ \nl
SE source & & \nl
0.5--2  & $0.27\pm0.08\times10^{-14}$ & $0.39\pm0.11\times10^{40}$ \nl
           
\enddata

\end{deluxetable}

\begin{figure*}[htp]
\centerline{\includegraphics[angle=0,width=5in]{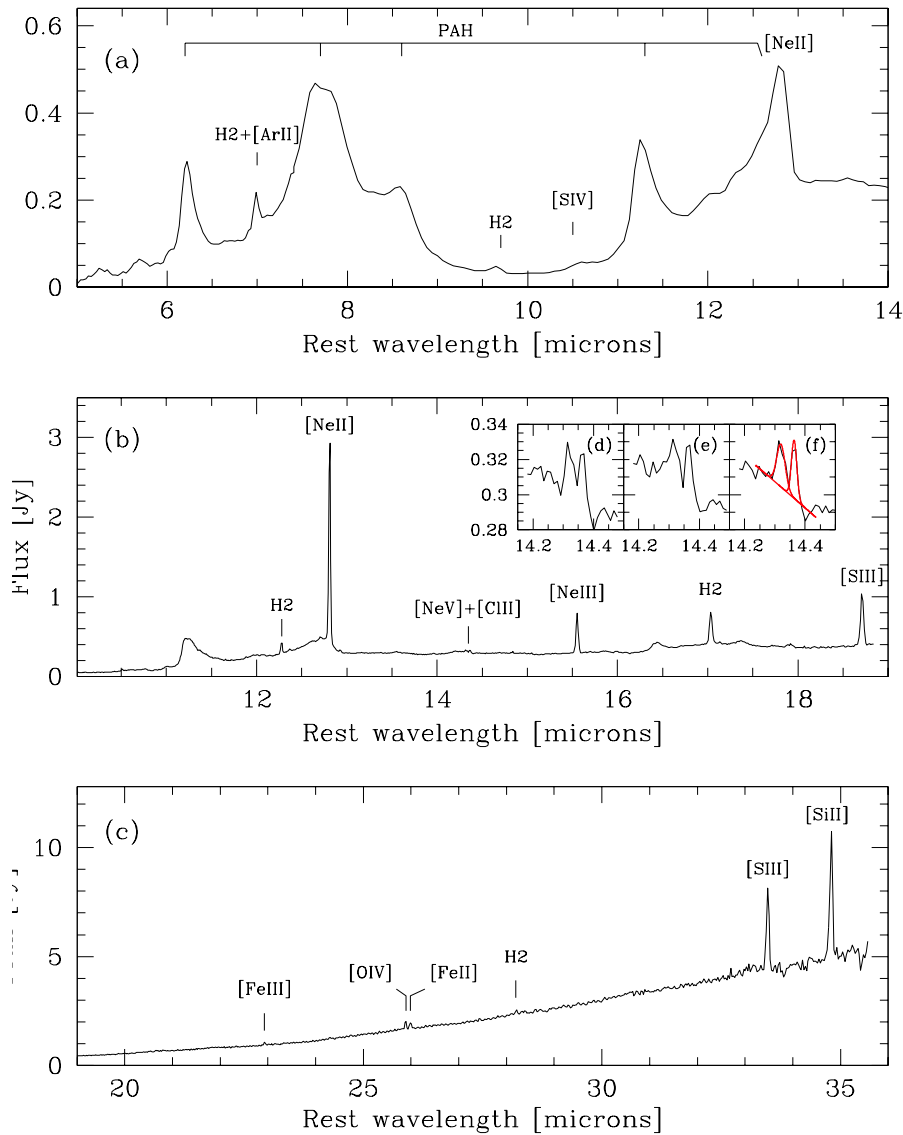}}
\figcaption{Low (a) and high (b and c) resolution {\it Spitzer} IRS
spectrum of IC 883. The insets in (b), which are labeled (d, e, and f), are
centered on the location of the [Ne V] 14.3$\mu$m and [Cl II] 14.4$\mu$m
lines. Insets (d) and (e) are spectra from the two individual nods, and (f)
is the average of the two nods. The simultaneous fit to the [Ne V] and [Cl
II] emission features are shown in red (3f). The narrowness of the [Cl II] line
is likely due to the trough observed in both nods at 14.4$\mu$m.} 
\end{figure*}

The low+high resolution composite {\it Spitzer} IRS spectrum (Figure 4)
shows strong emission from forbidden lines, H$_2$ emission, and PAH. The
[Ne II] 12.8$\mu$m, [Ne III] 15.6$\mu$m, H$_2$ $S(1)$, [S III] 33.48$\mu$m
and [Si II] 34.82$\mu$m emission lines are the strongest non-PAH emission
features in the spectrum. The fluxes of the detected forbidden and H$_2$
emission lines are provided in Table 2. Further, the equivalent width of
the 6.4 $\mu$m PAH emission feature (Figure 4a) is measured to be 0.39
$\mu$m.

\begin{deluxetable}{lr}
\tablenum{2}
\tablewidth{0pt}
\tablecaption{IRS Emission Lines and Line Flux Densities}
\tablehead{
\multicolumn{1}{c}{Emission Line} &
\multicolumn{1}{c}{Flux Density}
\nl
\multicolumn{1}{c}{} &
\multicolumn{1}{c}{(in W cm$^{-2}$)}
}
\startdata

H$_2$ S(3) 9.66$\mu$m & $6.52\pm0.41\times10^{-21}$ \nl
$[$S IV$]$ 10.51$\mu$m & $1.20\pm0.14\times10^{-21}$ \nl
H$_2$ S(2) 12.28$\mu$m & $6.15\pm0.32\times10^{-21}$ \nl
Hu-$\alpha$ 12.37$\mu$m & $1.04\pm0.11\times10^{-21}$ \nl
$[$Ne II$]$ 12.8$\mu$m & $1.13{\pm0.012}\times10^{-19}$  \nl
$[$Ne V$]$ 14.32$\mu$m & $1.17{\pm0.23}\times10^{-21}$  \nl
$[$Ne III$]$ 15.55$\mu$m & $1.85{\pm0.018}\times10^{-20}$  \nl
H$_2$ S(1) 17.03$\mu$m & $1.48{\pm0.037}\times10^{-20}$  \nl
$[$S III$]$ 18.71$\mu$m & $2.20{\pm0.061}\times10^{-20}$  \nl
$[$Fe III$]$ 22.92$\mu$m & $2.10\pm0.11\times10^{-21}$ \nl
$[$O IV$]$ 25.89$\mu$m & $6.53\pm0.072\times10^{-21}$ \nl
$[$Fe II$]$ 25.99$\mu$m & $5.53\pm0.67\times10^{-21}$ \nl
H$_2$ S(0) 28.22$\mu$m & $2.47\pm0.49\times10^{-21}$ \nl
$[$S III$]$ 33.48$\mu$m & $6.41\pm0.39\times10^{-20}$\nl
$[$Si II$]$ 34.82$\mu$m & $9.77\pm0.72\times10^{-20}$ \nl

\enddata

\end{deluxetable}

\section{Star Cluster Identification and Photometry}

A major goal of the present paper is to perform a photometric analysis of
the luminous star clusters visible in the {\it HST} images. Before
an automated routine for cluster identification could be applied to the
images, contamination from foreground stars and distant background galaxies
in the images had to be minimized. To address this, a mask of IC 883 was made
by first creating a 21$\times$21 median-smoothed version of the F435W and
F814W images. The effect of this filtering is to minimize structures with
spatial extents significantly smaller than the filter size (i.e., faint
stars and distant, background galaxies). The image was then boxcar smoothed
(40$\times$40). The result of the filtering and smoothing is an image with
high valued pixels where the galaxy is located (i.e., where the median
filtering would produce a pixel value greater than zero) and low valued
pixels in the background regions (where the median filtering would negate
pixel counts associated with distant, background galaxies and faint
foreground stars).  The backgrounds, containing low pixel values, were then
set to zero, while the high pixels were set to one. Finally, pixels
associated with any bright stars in the image were set to zero. The
resultant mask, which outlines the galaxy to a surface brightness of $\sim
25$ mag/${\arcsec}^2$, was then multiplied by the original reduced image of
the galaxy to set the regions outside of the galaxy equal to zero. The main
body of IC 883 in the masked F435W image is shown in Figure 5a.

\begin{figure*}[htp]
\centerline{\includegraphics[angle=-90,width=5in]{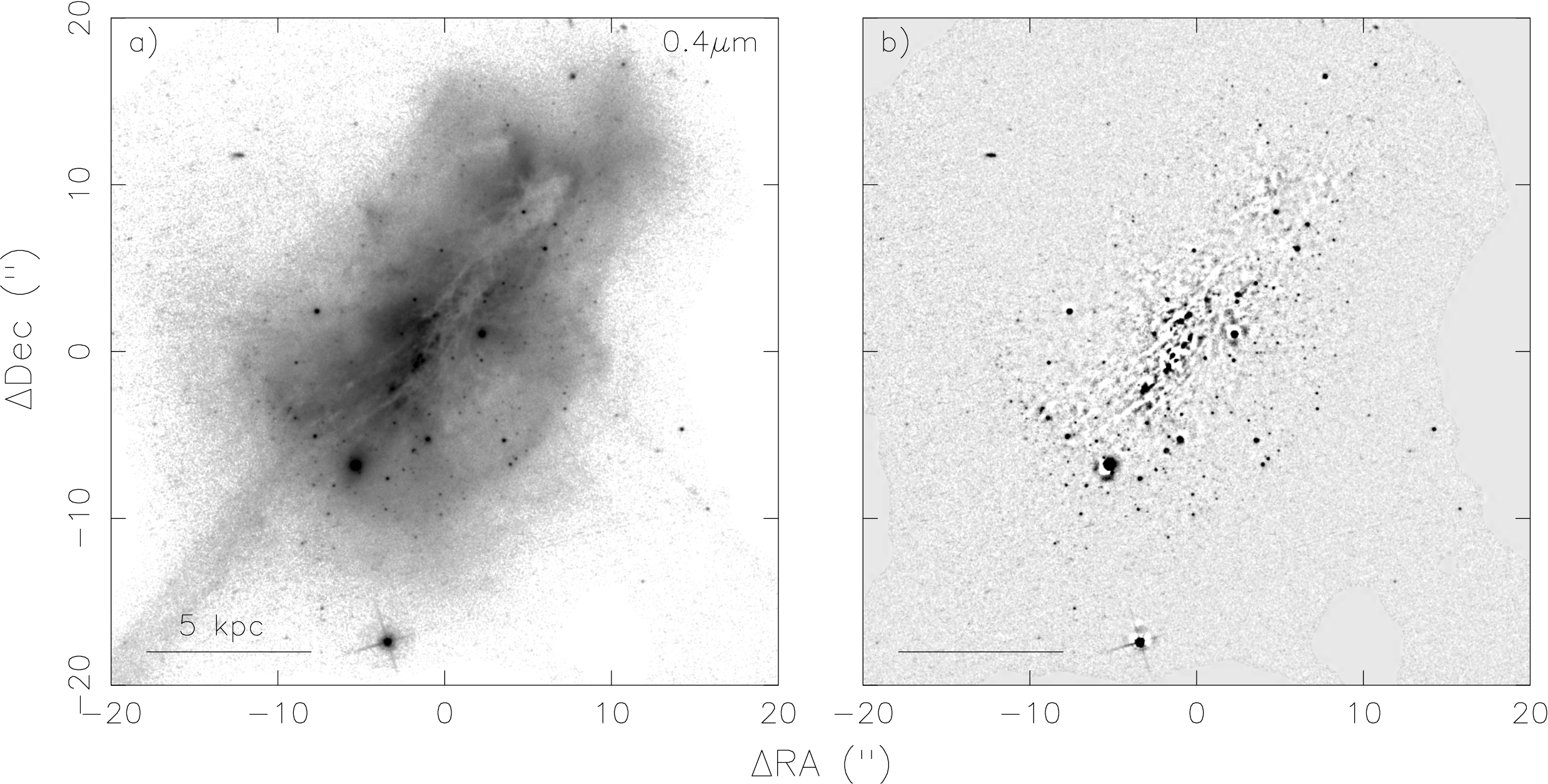}}
\figcaption{The {\it HST} F435W images of IC 883. (a) The original image of
the galaxy.  A mask has been applied to the image (see \S4).  (b) The image
of the galaxy after it has been processed by SOURCE EXTRACTOR. Note that the
residual artifacts in the nuclear region of IC 883 are filtered out by the
cluster identification criteria. In both images, North is up, East is to
the left.} 
\end{figure*}

Once the mask was applied, the clusters in IC 883 were identified using
SOURCE EXTRACTOR (Bertin \& Arnouts 1996). The
identification of clusters and the extraction of photometry is complicated
by the non-uniform surface brightness of the underlying galaxy.  To
estimate and subtract the underlying galaxy, SOURCE EXTRACTOR 
iteratively computes the median and standard
deviation of the pixels within a mesh of $n\times n$ pixels; during each iteration, 
outlier pixels are discarded until 
all of the pixels within each mesh are
within $\pm3\sigma$ of the median
value. Several mesh sizes were tested; for each mesh, the photometry of several of the clusters 
were computed ``by hand'' via the image display and analysis program 
IPAC SKYVIEW (http://www.ipac.caltech.edu/skyview/) and compared to values estimated from the 
original image. The mesh size of $9\times9$ pixels did an efficient job of
removing the galaxy and minimizing the creation of holes surrounding
clusters during the extraction, and thus preserving the integrity of the cluster photometry. 
The F435W image with the mesh applied is
showed in Figure 5b. 

After SOURCE EXTRACTOR was used to fit the underlying galaxy and to identify
candidate clusters, the photometry of the candidates was computed using IDL routines.
The photometry was calculated using the latest Vega magnitude
zero-points for the ACS/WFC F435W and F814W filters tabulated at
http://www.stsci.edu/hst/acs/analysis/zeropoints.
Photometric aperture corrections were applied to the
0.3$^{\arcsec}$ diameter aperture measurements using determinations of the
curve of growth of bright, unresolved stars and clusters in the GOALS sample images; these curve
of growth measurements are in agreement with values measured by Sirianni et
al. (2005: Table 3).  The final selection of clusters was based on the
following criteria: (1) a detection at a signal-to-noise ratio $S/N \geq 5.0\sigma$ in both the F435W and F814W
images and (2) a
full-width at half the maximum (FWHM) intensity in the range of $\sim 1.7 -
4.0$ pixels, which is consistent with the clusters being unresolved or
partially resolved (4.0 pixels $\sim 100$ pc at the redshift of IC 883). 
Only cluster candidates which meet these criteria are used in the analysis
that follows. Using this method, 156 star clusters are detected
in IC 883.

The completeness limits were tested using the method described in detail in
Vavilkin et al. (2011). For the IC 883 images, the F435W and F814W images
are found to be complete at the 50\% level at the apparent magnitude
$m_{\rm F435W} \sim 26.1$ (absolute magnitude of $M_{\rm F435W} \sim
-9.15$) and $m_{\rm F814W} \sim 25.63$ ($M_{\rm F814W} \sim -9.9$) mags,
respectively.

\begin{figure*}[htp]
\centerline{\includegraphics[angle=-90,width=5in]{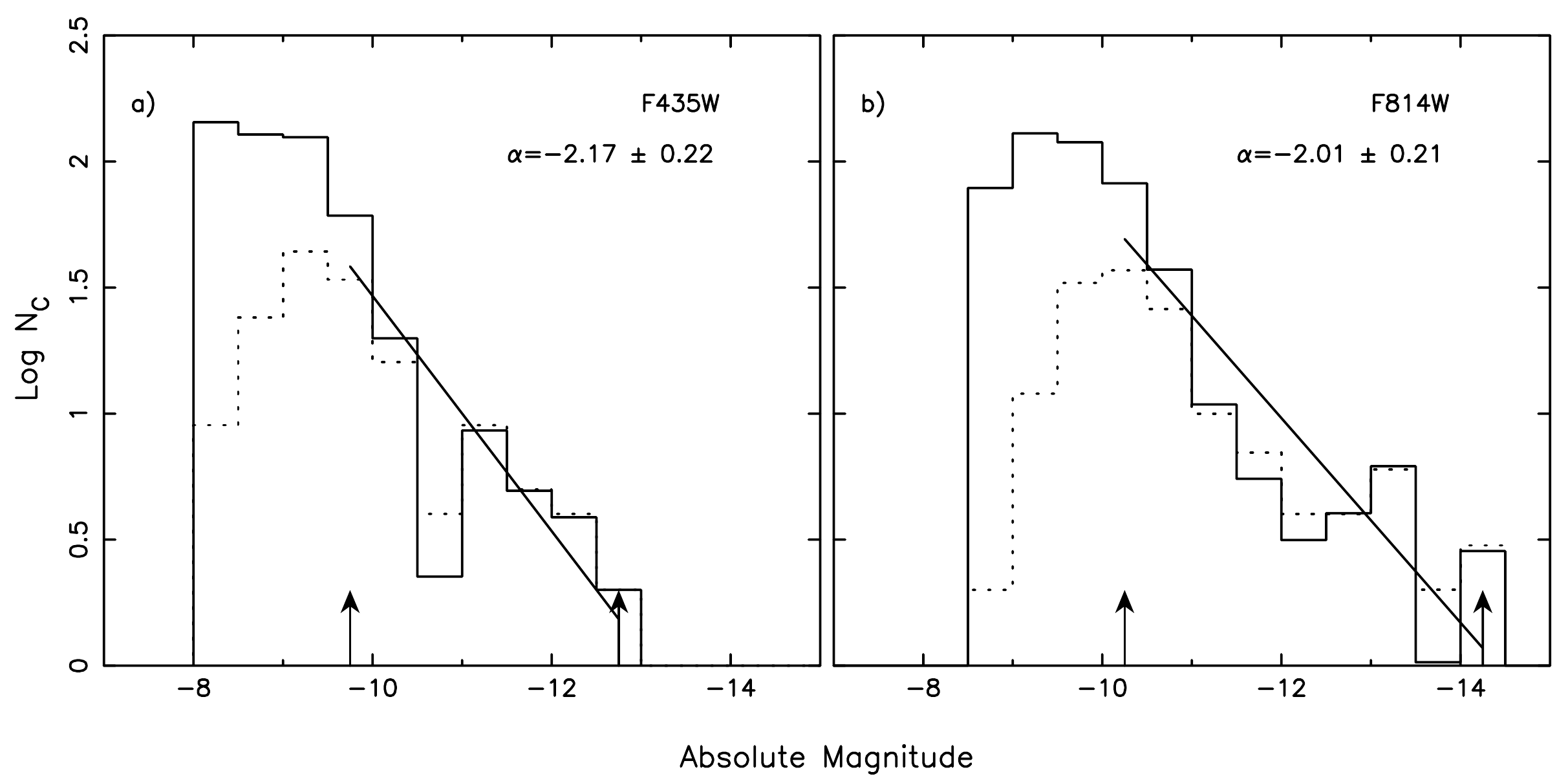}}
\figcaption{The luminosity function for clusters visible in the F435W (a)
and F814W (b) images of IC 883.  The dashed line histogram is the raw,
uncorrected luminosity function, whereas the solid line histogram
luminosity function has been corrected for foreground stars and background
galaxy contamination, and for completeness. The solid line represents a
chi-squared fit to the histogram bins over the magnitude ranges of $M_{\rm
F435W} -9$ to $-12$ mag and $M_{\rm F814W} \sim -10$ to $-13$ mag. The
arrows mark the range of the fits, which
are done to the 50\% completeness level.  }
\end{figure*}

As stated above, the regions around the galaxy were masked before the
clusters were identified, extracted and analyzed. In order to estimate the
contamination in each image by foreground stars and distant galaxies,
these same procedures were also run on galaxy masked,  ``sky-only''
versions of the ACS images. The ratio of the area of the image subtended by
the galaxy relative to the area subtended by ``sky-only'' provides a
normalization to the sky-only source counts as a function of magnitude,
and thus a measure of source contamination toward the galaxy.

\section{Discussion}

\subsection{Properties of the Optically-Visible Star Clusters}

One hundred fifty-six optical star clusters have been detected in both the
nuclear region, main body and the extended tidal tails of IC 883.  At the
resolution of the F435W images, the best constraint on the sizes of the
unresolved star clusters is less than $\sim 57$ pc in diameter. Thus,
regardless of the high-resolution nature of the images, the resolution is
significantly poorer than the median effective diameter of of Milky Way
globular clusters ($D_{\rm eff} \sim 6$ pc: van den Bergh 1996) and star
clusters observed in the Antennae Galaxy ($D_{\rm eff} \sim 8\pm2$ pc:
Whitmore et al. 1999).

Figure 6 is a histogram plot of the logarithm of the number of star
clusters versus the cluster absolute magnitude, $\log(N)$ versus $M$, as
measured in both the F435W and F814W images.  Both the raw histogram and
the histogram corrected for background contamination and completeness are
shown, with the bulk of the clusters observed to have $M_{\rm F814W} \sim
-13$ to $-9$ mag.  A common parameter derived from such a plot is the power
law slope, $\alpha$, of the cluster luminosity function. I.e., the cluster
luminosity function, $\psi (L)$, expressed in terms of luminosity, $L$, is
of the form

$$dN = \psi (L) dL \propto L^\alpha dL, \eqno(1)$$

\noindent
where $dN$ is the number of clusters between $L$ and $L+dL$.
The
best fit luminosity function over the $M_{\rm F814W} \sim -13$ to $-10$
magnitude range (i.e., down to a completeness level of 50\%) has an
$\alpha_{\rm F814W} = -2.01\pm0.21$, where $\alpha$ is related to the
slope, $a$ (= $d \log N / d M$), of the histogram plotted in Figure 6 via

$$\alpha = -(2a +1). \eqno(2)$$

\noindent
For the F435W data, the best fit function over $M_{\rm F435W} \sim -12$ to
$-9$ mag  is $\alpha_{\rm F435W} = -2.17\pm0.22$. While a direct comparison
cannot be made between the ``B"- and ``I"-band derived values for IC 883
and the $V$-band derived $\alpha$ values for other well-studied starburst
mergers, it is worth noting that the $\alpha$ values of IC 883 are, within
the errors, consistent with $\alpha$ derived at $V$-band for NGC 0034
($\sim -1.73\pm0.1$: Schweizer \& Seitzer 2007) and the Antenna Galaxy
($\sim -2.13\pm0.07$: Whitmore et al. 2010), but higher than the $\alpha$ of
ESO 350-IG038 ($\sim -1.52\pm0.05$: Adamo et al.  2010).

\begin{figure*}
\centerline{\includegraphics[angle=-90,width=5in]{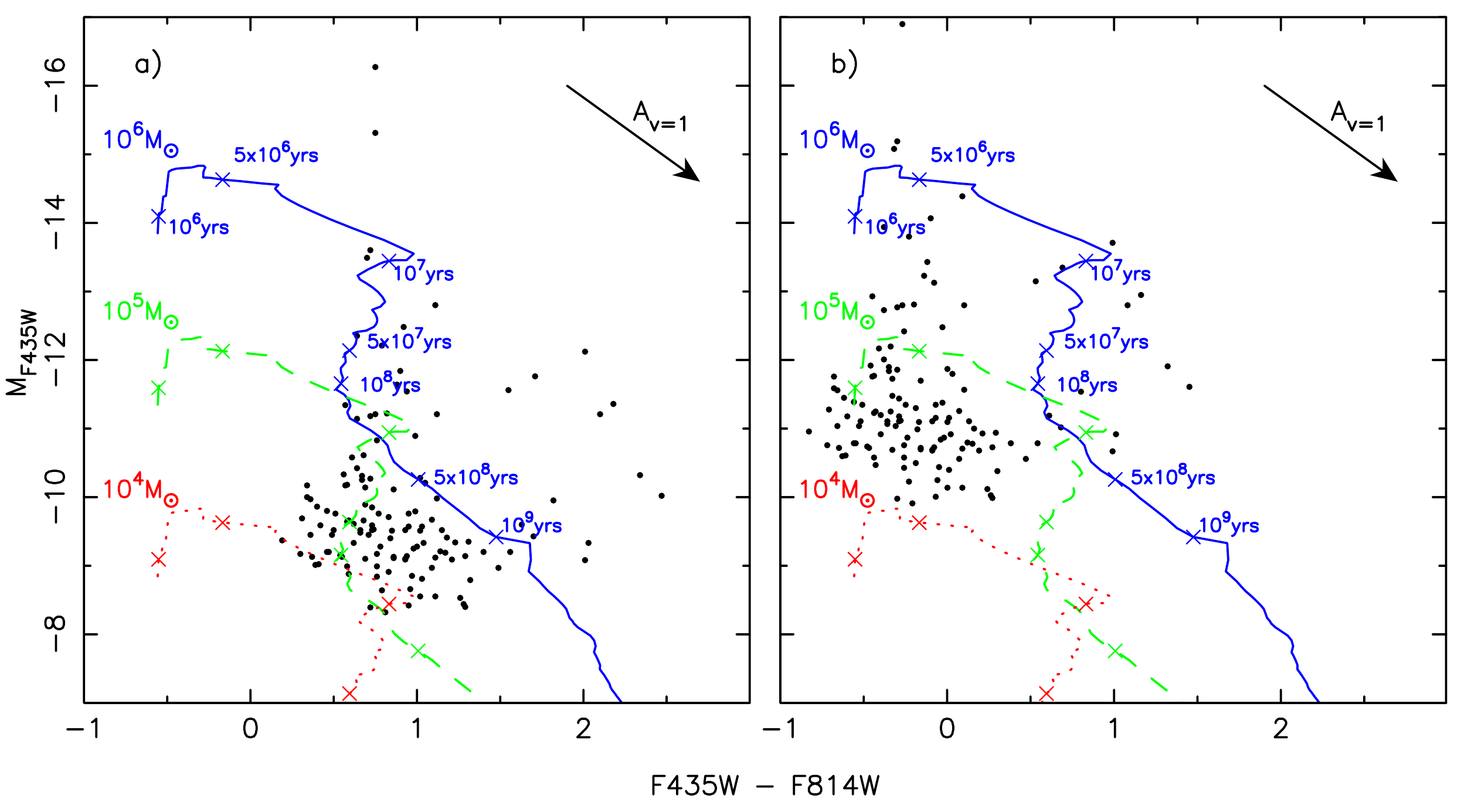}}
\figcaption{(a) The $F435W-F814W$ versus $M_{\rm F435W}$ color-magnitude
diagram. Bruzual Charlot population synthesis models with instantaneous
starburst, solar metallicity are plotted for cluster masses of $10^4$, $10^5$ and
$10^6$ M$_\odot$. The plotted cluster data have not been corrected for
reddening. Only clusters with color uncertainties $< 0.2$ mags are plotted. 
The vector represents the magnitude and direction of a cluster in the diagram
having one magnitude of visual extinction.
For the extinction measurements, the Cardelli et al. (1989)
extinction model was used.
(b) Same as (a), except the clusters have been corrected for an
extinction, $A_V = 1.2$. 
} 
\end{figure*}

\begin{figure*}
\centerline{\includegraphics[angle=-90,width=5in]{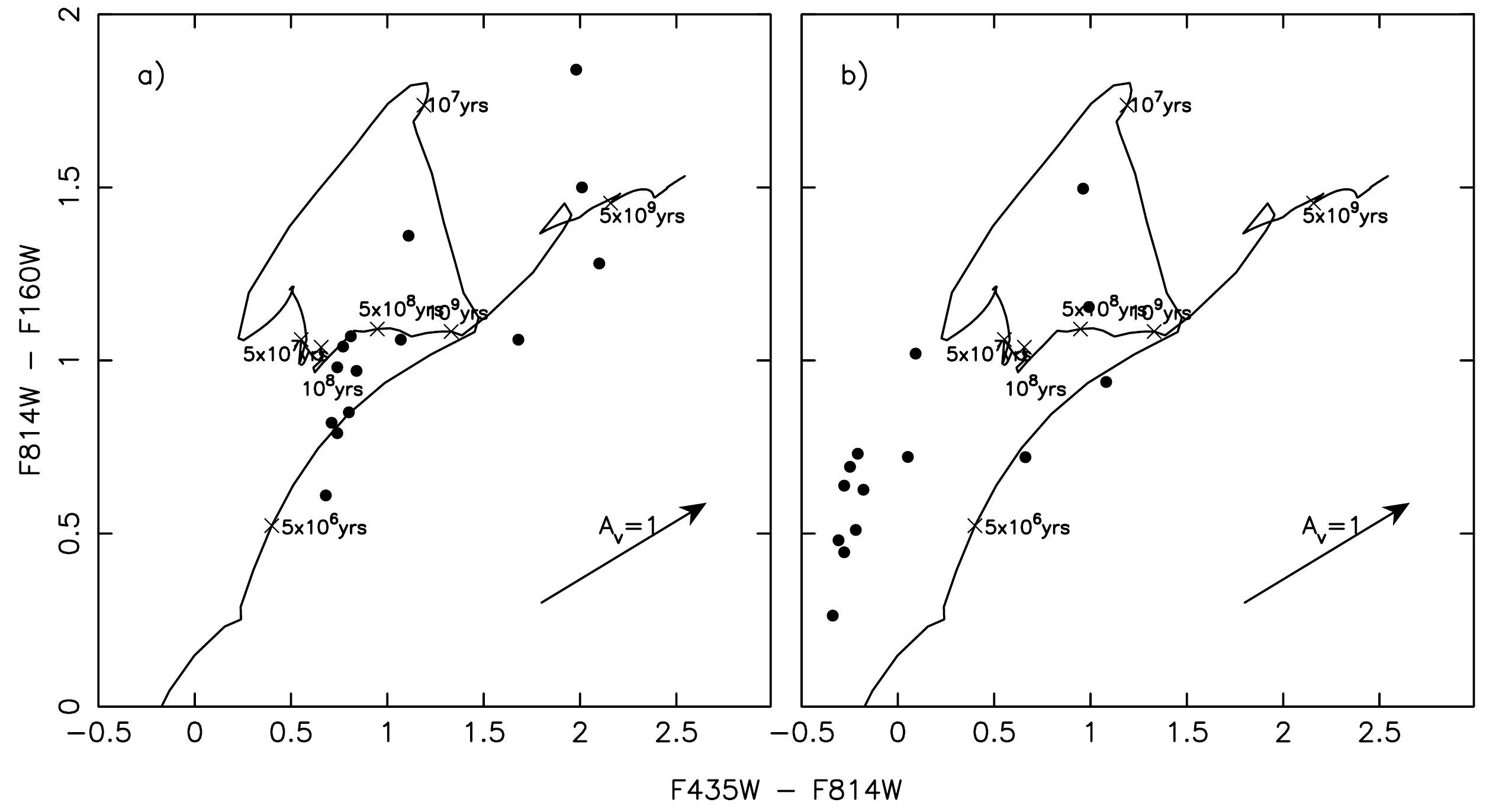}}
\figcaption{(a)The $F435W-F814W$ versus $F814W-F160W$ color-color diagram.
Bruzual Charlot population synthesis models with instantaneous starburst,
solar metallicity are plotted. The plotted cluster data have not been
corrected for reddening. The color uncertainties of the plotted clusters are $< 0.13$ mags.
The vector represents the magnitude and direction of a cluster in the diagram
having one magnitude of visual extinction. For the extinction measurements, the Cardelli 
extinction model was used.
(b) Same as (a), except the clusters have been corrected for an
extinction, $A_V = 1.2$. 
} 
\end{figure*}

The $F435W-F814W$ versus $M_{\rm F435W}$ 
color-magnitude diagram is show in
Figure 7. The approximate range of cluster ages is estimated from the data
in Figure 7a by assuming a Bruzual Charlot (Bruzual \& Charlot 2003)
population synthesis model with an instantaneous starburst, solar metallicity,
and a cluster mass of $10^5$ M$_\odot$ (e.g., see Figure 8 of Surace et al.
1998). Models of clusters with masses of $10^4$ and $10^6$
M$_\odot$ is also shown for comparison. The majority of the clusters lie
within the few$\times10^7 - 10^8$ year age range; there is also overlap
with an age range of $5-10\times10^6$ years for $10^4$ M$_\odot$ clusters and 
$> 10^8$ years for $10^6$
M$_\odot$ clusters. By comparison, a similar
analysis has been done by making use of archival $1.6\mu$m (F160W filter)
NICMOS data of IC 883 (see
Scoville et al. 2000). The ACS--NICMOS color-color diagram ($F435W-F814W$)
vs. ($F814W-F160W$), which contains the small number of clusters bright
enough to be detectable in 
the inner $19.5\arcsec\times19.2\arcsec$ field of view of NIC2 centered on the nucleus,
is shown in Figure 8a.
If a synthesis model with identical parameters as that applied to the 
ACS data are used, the ages
for the majority of the clusters in the range of $\sim 5-500\times10^6$ years. 
These ages are, of course, highly uncertain due to the unknown affects
of dust, and the unknown cluster masses, and metallicities of the clusters.

While measurements of the extinction in the main body of IC 883 do not
exist, Veilleux et al. (1995) made long-slit measurements of the $E (B -
V)$ of several LIRGs as part of the original Bright Galaxy Survey. For
galaxies with H$\alpha$ and H$\beta$ measurements at a nuclear distance $ >
3$ kpc -- i.e., where the bulk of the clusters in IC 883 reside (e.g.,
see \S 5.2) -- and where the extinction in the long slit measurements
is decreasing with radius, the $E
(B - V) \sim 0.4 - 0.7$. This corresponds to a visual extinction of $A_V
\sim 1.2 - 2.2$.  This is also consistent with a multi-wavelength analysis of clusters in
the starburst galaxy ESO 350-IG038 (Adamo et al. 2010), in which the authors derive
extinction values toward the majority of the clusters of $A_V \sim 0 - 2.2$. 
If an $A_V = 1.2$ is adopted for the clusters in IC
883, the age estimates  (Figure 7b) for the bulk of the clusters range from $10^6$ to several
$\times10^7$ years for $\sim 10^5$ M$_\odot$ clusters. In Figure 8b, such large
extinction values do not fit the solar metallicity models for the bulk of the clusters,
but are consistent with ages ranging from $10^{6-8}$ years for the reddest clusters.

\subsection{The Distribution of Star Clusters Relative to
Emission at Other Wavelengths}

\begin{figure*}[htp]
\centerline{\includegraphics[angle=0,width=5in]{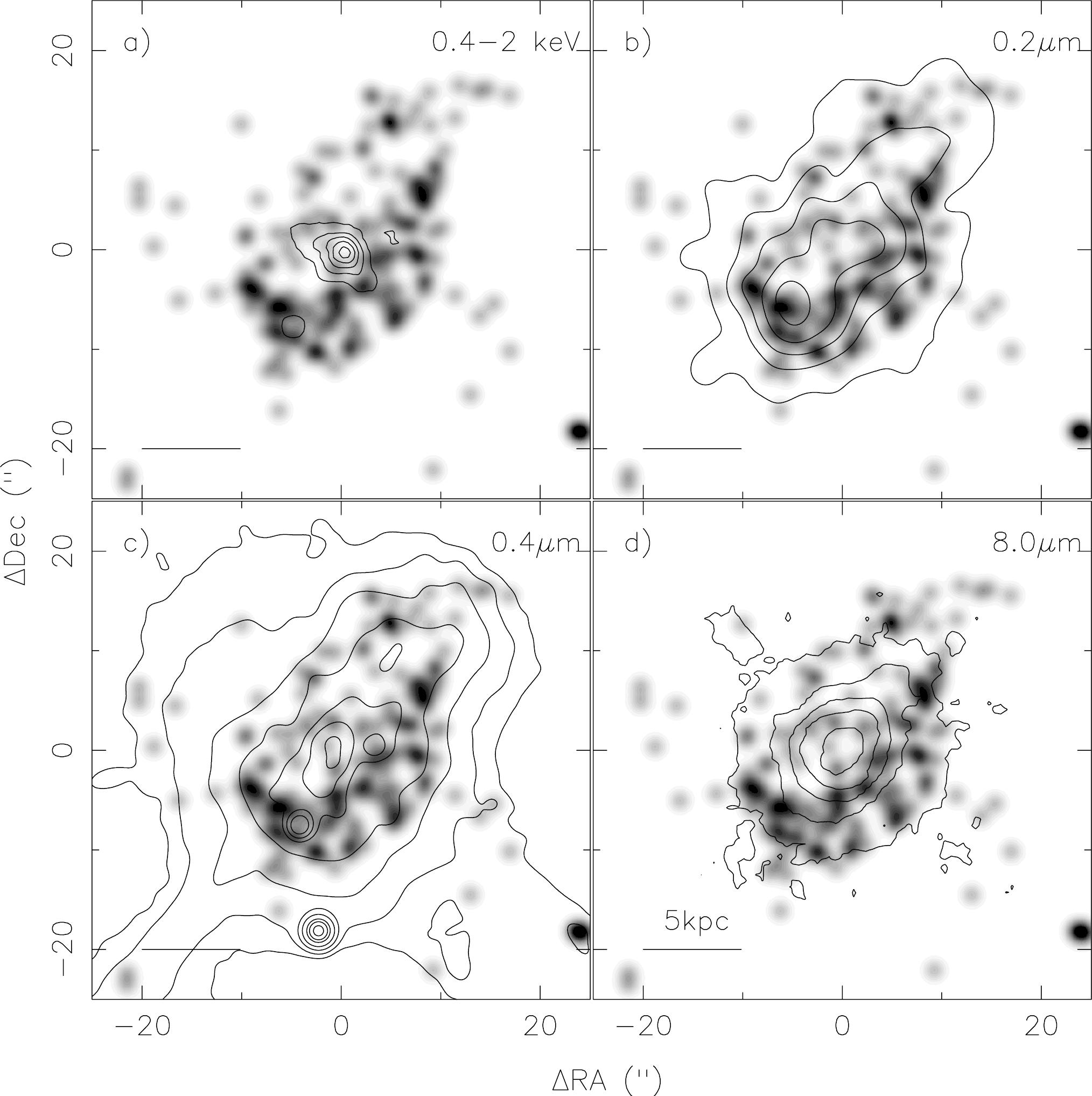}}
\figcaption{A grey scale image of star cluster density versus contours of
the (a) Chandra X-ray, (b) GALEX $0.23\mu$m, (c) HST 0.4$\mu$m, and (d)
IRAC 8$\mu$m. In all images, North is up, and East is to the left.}
\end{figure*}

Figure 9 shows the optically-visible star cluster density in grayscale
relative to the underlying galaxy (in contours) in the soft X-ray band, and
at  0.23$\mu$m, 0.44$\mu$m, and 8$\mu$m. The star cluster density has been
derived by first creating an image containing the position of each star
cluster (each with intensity of unity), then applying gaussian smoothing to
the image with a $\sigma = 20$ pixels, i.e., the resolution of the 8$\mu$m
IRAC image.  While there are clusters present throughout most of the main
body of IC 883, the highest cluster density regions are located along an
arc in the eastern and southern region of the main body. Clusters are
relatively absent in projection towards the optical, X-ray and 8$\mu$m
cores. In contrast, the peak of the 0.23$\mu$m emission corresponds to the
clusters observed in the SE portion of the arc. The offset between the
8$\mu$m cores and the cluster arcs is also illustrated in the histogram in
Figure 10, which shows the number of clusters as a function of the distance
from the 8$\mu$m core. The majority of the clusters are at a distance of
3--7 kpc from the core -- the peak of the 0.23$\mu$m emission corresponds
to a distance of 3.2 kpc from the core.

\begin{figure}
\centerline{\includegraphics[angle=-90,width=3in]{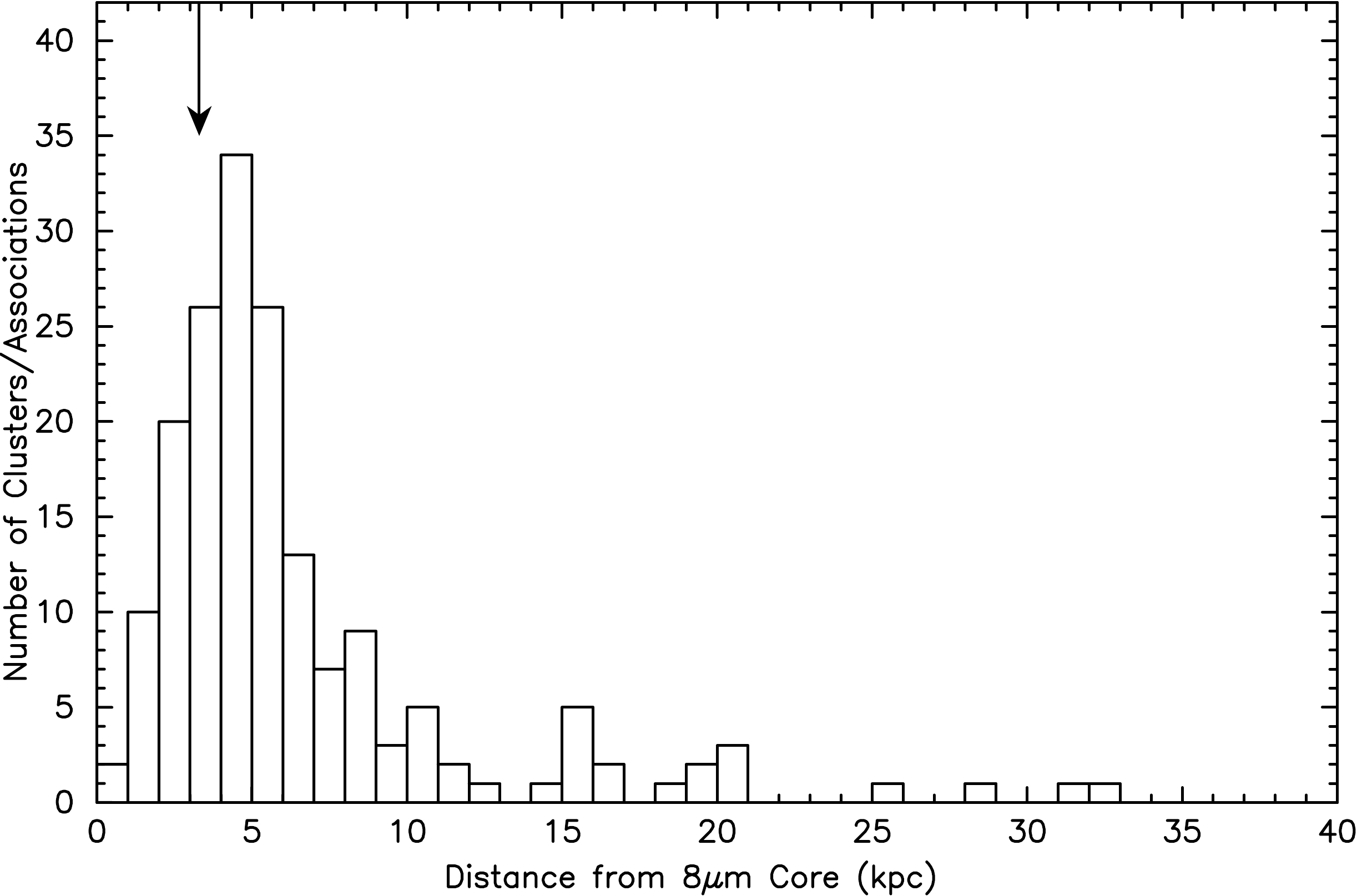}}
\figcaption{The number of star clusters versus their
distance from the IRAC 8$\mu$m core.  The position of the GALEX 0.23$\mu$m
peak is indicated on the plot as an arrow and corresponds to a distance of
3.2 kpc from the 8$\mu$m core.} 
\end{figure}

\begin{figure}
\centerline{\includegraphics[angle=0,width=3in]{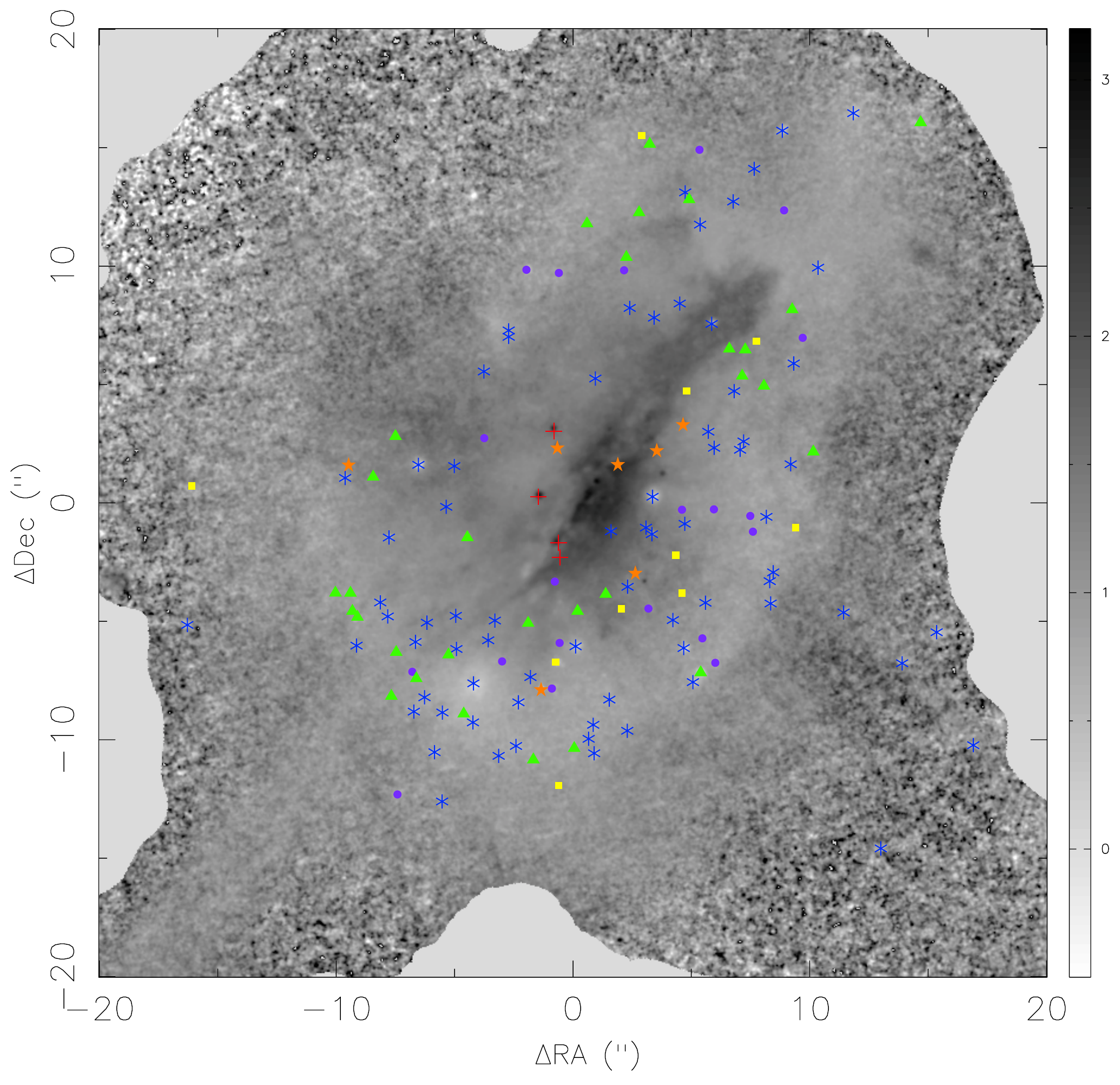}}
\figcaption{A plot of the position of clusters superposed on a
$F435W-F814W$ greyscale image of IC 883.  The color value of the cluster
represents the $F435W-F814W$ bin of the cluster. Only clusters with
$F435W-F814W$ mag error less than 0.2 are plotted. The $F435W-F814W = 0 -
0.5$ bin clusters are designated with violet circles.  $F435W-F814W =
0.5-1$ are designated with blue asterisk. The $F435W-F814W = 1-1.5$ are
designated with green triangles.  The $F435W-F814W = 1.5 - 2$ are
designated with yellow squares.  The $F435W-F814W = 2 - 2.5$ are designated
orange stars.  $F435W-F814W > 2.5$ are designated with red crosses.}
\end{figure}

The lack of star clusters observed toward the core of the merger is undoubtedly the
result of heavy dust extinction. In Figure 11, the cluster positions are
superposed on the F814W band image of IC 883. Each cluster is coded with a
color which represents the $F435W - F814W$ color bin the cluster belongs to
(see the Figure 11 caption). As is clear from the Figure, the majority of
the reddest clusters lie in or near the reddest 
nuclear regions of the galaxy. Further
support for nuclear reddening comes from the fact that two of the nuclear
clusters visible in the F814W image are not visible in the F435W image, and
one of the nuclear clusters visible in the NICMOS F160W image is not
visible in the F435W image.

The ``redness'' of the nuclear region is also illustrated in Figure 12a--c,
which shows the $F435W - F814W$ greyscale image of IC 883, with contours of
the X-ray, NUV, and 8$\mu$m contours superimposed. Figure 12d shows the
1.6$\mu$m HST NICMOS image in greyscale with the VLA radio, CO and 8$\mu$m
contours superposed. The reddest portion of IC 883 clearly lies in a linear
strip; in the 1.6$\mu$m image, the high surface brightness region of the
strip is $\sim 2$ kpc across, and the red region in the $F435W - F814W$
image is $\sim 9$ kpc in extent. The core of the primary X-ray emission,
the bulk of the mid-IR emission, and all of the VLA and OVRO-detected radio
and CO emission, respectively, lie in these regions.

\begin{figure*}[htp]
\centerline{\includegraphics[angle=0,width=5in]{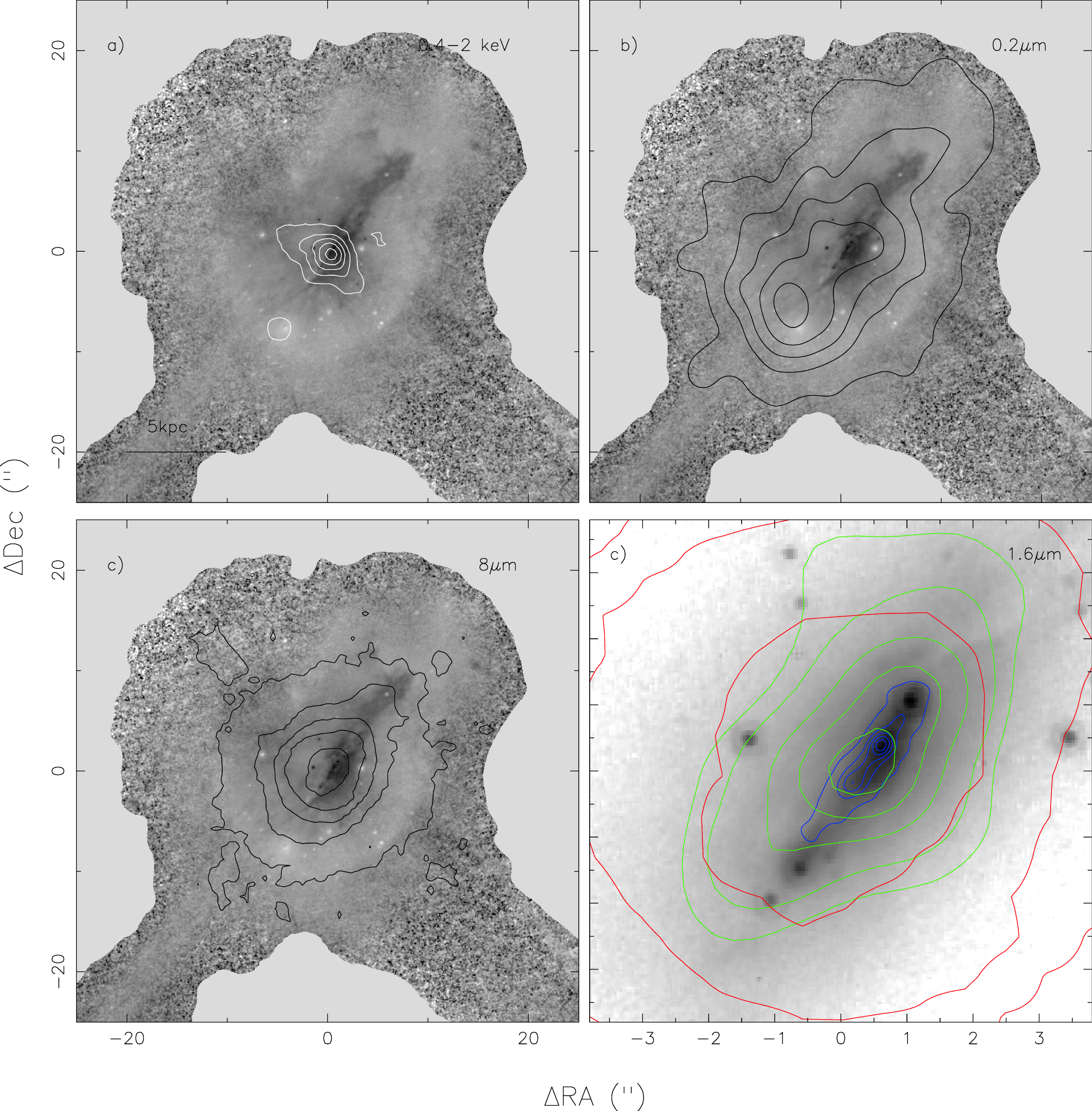}}
\figcaption{A greyscale of $F435W - F814W$ versus contours of (a) Chandra
X-ray, (b) GALEX $0.23\mu$m, (c) and IRAC 8$\mu$m. (d) Greyscale 1.6$\mu$m
HST NICMOS image with contours of VLA 1.5 GHz (blue), CO(1$\to$0) (green)
and IRAC 8$\mu$m (red) superposed. The F435W--F814W greyscale intensity values
are the same as those plotted in Figure 11. In all images, North is up, and East is
to the left.} 
\end{figure*}

\subsection{Nuclear Activity and Environment}

Both the {\it Chandra} X-ray and {\it Spitzer} IRS observations provide
probes into the extinguished nuclear regions of IC 883. The hard X-ray
emission is extended, and the X-ray spectrum (Figure 3) has a hardness
ratio which is consistent with X-ray emission from a starburst population.
Further, the observed $L_{\rm X(2-10~keV)} / L_{\rm FIR(40-120\mu m)} \sim
8\times10^{-5}$, is also consistent with the ratio observed for lower
luminosity, H II region-dominated galaxies ($L_X / L_{\rm FIR} \sim
2\times10^{-4}$: Ranalli et al. 2003). Finally, the large equivalent width
of the Si XIII line may be due to a contribution of AGN photoionized gas
(\S 3; Iwasawa et al. 2011).

The IRS data (Figure 4) provide further support for the X-ray data
interpretation of a starburst dominated galaxy with some evidence of an AGN: the
low equivalent width of the 6.2$\mu$m PAH emission feature ($\sim
0.39\mu$m) is consistent with those measured for AGN hosts. Further, the 8$\sigma$
detection of the high ionization [Ne V] 14.32$\mu$m line is additional
evidence that an AGN may be present in IC 883. The [Ne V] / [Ne II]
12.8$\mu$m flux ratio of $\sim 0.01$, as well as the low [O IV] 25.9$\mu$m
/ [Ne II] 12.8$\mu$m ratio ($\sim 0.06$), are indications that the AGN is
either extremely weak, or that the AGN (and thus the [Ne V] emitting
region) is heavily enshrouded in dust relative to the region(s) where most
of the [Ne II] emission is being generated (e.g., see discussion of
mid-infrared line diagnostics in Armus et al. 2007; Veilleux et al. 2009).
However, the lack of a strong 9.7$\mu$m silicate absorption feature at 10$\mu$m
rules out any direct evidence of the second possibility.

Is the detection of the [Ne V] line truly an indication that an AGN is
present? A possible way in which the line might be generated by stellar
processes is via supernovae. Consider the supernova remnants (SNRs) 1E0102-72.3 and RCW103:
Rho et al. (2009) and Oliva et al. (1999) published a mid-infrared spectrum of 1E0102-72.3 and 
RCW103, respectively, which show
them to have [Ne V] / [Ne II] $\sim 0.07$ and $0.01$, respectively. Both measurements
were made in small apertures ($3.7\arcsec\times57\arcsec$ for 1E0102-72.3 and $14\arcsec\times27\arcsec$ for RCW103). 
Thus, if the conservative assumption is made that the [Ne V] emission
measured in the small apertures is uniformly distributed over the SNRs (i.e., $44\arcsec$ diameter region for 1E0102-72.3
and $105\arcsec\times60\arcsec$ region for RCW103), the [Ne V]
luminosities of 1E0102-72.3 and RCW103 are $L_{\rm [Ne V]} {\rm [SNR]} \sim
5.8\times10^{27}$ W and $2.5\times10^{26}$ W, respectively.  If it is further assumed these SNRs are
representative of SNRs in IC 883 and
that the measured [Ne V] flux of IC 883 is its intrinsic flux, approximately $0.3-7\times10^6$ SNRs
are required to produce the luminosity of the [Ne V] line observed in IC
883 ($L_{\rm [Ne V]} {\rm [IC883]} = 1.69\times10^{33}$ W). If these
[Ne V]-emitting remnants last $\sim1000$ years on average, IC 883 would require a steady state
production of $0.3-7\times10^3$ SNRs yr$^{-1}$ to sustain the observed [Ne V]
luminosity of IC 883. This SNR rate can be compared directly with the rate expected for IC 883 -- 
the star formation rate of IC 883, as measured from
either the infrared or radio luminosity and using prescriptions in
Kennicutt et al. (1998) and Bell (2003), respectively, is $\dot{M}_* \sim 80$
M$_\odot$ yr$^{-1}$. For a Salpeter initial mass function (IMF), $\xi (M) \propto M^{-2.35}$ (Salpeter 1955), 
the
number of SNRs produced in IC 883 per year is

$$\dot{N}_{\rm SNR} \sim \dot{M}_* { 
\int^{\rm 120M_\odot}_{\rm 8M_\odot}  \xi (M) dM \over 
\int^{\rm 120M_\odot}_{\rm 0.5M_\odot} M \xi (M) dM
} 
\sim 1~{\rm SNR~yr^{-1}}. \eqno(3)$$

\noindent
Thus, even if the estimated number of SNRs required to
produce $L_{\rm [NeV]}$ (IC 883) is an order of magnitude off, the
required SNR rate would still be too high given the estimated star
formation rate.  

%It is also worth noting that, even given the above
%calculation, the majority of the [Ne II] emission in IC 883 is most likely
%associated with massive stars on the main sequence and not SNRs, and thus
%the [Ne V] / [Ne II] produced by any SNRs present would have to be
%significantly greater than that measured for RCW103 (i.e., [Ne V] / [Ne II]
%$>> 0.01$). This would rule out RCW103-type SNRs as the sole source of the
%[Ne V] emission.

The nuclear environment in which the nuclear starburst and weak AGN reside
can be assessed with the multi-wavelength imaging data presented in Figure
1. While the {\it Spitzer} 70$\mu$m MIPS image is of insufficient
resolution ($\sim 18\arcsec \sim 9$ kpc) to constrain the extent of the
far-IR emission, the extents of the high surface brightness {\it HST} near-IR
(Scoville et al. 2000; Figure 12d), ground-based mid-IR (Soifer et al. 2001; Figure 1),
radio, and CO emission (e.g., Figure 1) are similar. As a measure of the
extent of the far-IR emission, the radio emission will thus be used as a
proxy. This assumption seems reasonable, given that the radio emission is
tracing feedback from star formation and/or AGN activity; i.e., processes
which can heat their surrounding dust. The extent of the radio emission is
$D_{\rm 1.4GHz} \approx 2.0$ kpc, and it is inferred from the kinematics of
the CO data (Downes \& Solomon 1998) that the CO, and most likely the radio
emission (i.e., if the radio emission is associated with star formation),
are distributed in a highly inclined disk.  Given $D_{\rm 1.4GHz}$, and the
$L_{\rm IR}$ of IC 883, the IR surface brightness is estimated to be

$$\mu_{\rm IR} \approx {4 L_{\rm IR} \over \pi D^2_{\rm 1.4GHz}} \approx
2\times10^{11} {\rm ~L_\odot ~kpc^{-2}}. \eqno(4)$$

\noindent
This surface brightness is an order of magnitude less than the theoretical
blackbody limit -- i.e., if it is assumed that the IR emission is emanating
from a spherical blackbody with a luminosity and temperature equivalent to
IR luminosity and dust temperature of IC 883, where the dust temperature is
estimated from the {\it IRAS} 60 and 100$\mu$m flux densities,
$f_{60\mu{\rm m}} = 17.04$ Jy and $f_{100\mu{\rm m}} = 24.38$ Jy (Sanders
et al. 2003),  via

$$T_{\rm dust} \approx -(1+z) \left[{82 \over \ln (0.3f_{60\mu{\rm
m}}/f_{100\mu{\rm m}})} - 0.5 \right] \approx 54 {\rm ~K}, \eqno(5)$$

\noindent
the blackbody size of the IR emitting region is

$$D_{\rm BB} = 2 \left[ {L_{\rm IR} \over 4 \pi \sigma T^4_{\rm dust}}
\right] ^{0.5} \approx 375 {\rm ~kpc}, \eqno(6)$$

\noindent
where $\sigma$ is the Stefan-Boltzmann constant. The blackbody limit IR
surface brightness is thus $5\times10^{12}$ L$_\odot$ kpc$^{-2}$. More
importantly, the estimated IR surface brightness is less than that of the
core of the Orion star-forming region ($\sim 2\times10^{12}$ L$_\odot$
kpc$^{-2}$) and less than that of LIRGs known to harbor luminous AGN ($\sim
10^{13-15}$ L$_\odot$ kpc$^{-2}$: see Soifer et al. 2000 and Evans et al.
2003). Thus, the IR surface brightness and the properties of the radio and
CO emission are consistent with the above assertion that the IR emission is
primarily generated by dust heated by star formation, and that the IR
starburst is 2 kpc in extent. As stated above, the estimated star formation
rate, calculated from either the IR (Kennicutt 1998) or radio (Bell 2003)
luminosity, is $\sim 80$ M$_\odot$ yr$^{-1}$. This nuclear starburst may
be driving the 0.5 -- 2 keV X-ray emission; the morphology of
the emission, which extends outward from 
the nucleus in a direction perpendicular to the radio 
and CO disk, is suggestive of a stellar
wind. Note however, that this possibility cannot be definitively concluded with the
present available data.

\subsection{The Southeastern X-ray Source}

As stated in \S 3, the X-ray emission has a SE component. This component is
coincident with a bright, compact optical knot in the optical images of IC
883. The knot is moderately resolved, with a FWHM extent of $\sim 55$ and
$63$ pc at F435W and F814W, respectively. The radial profile has a broader
shoulder than a gaussian, with an apparent full extent of $\sim 400$ pc.
Photometrically, it has a $M_{\rm F435W} = -16.58$ mag and a $F435W - F814W
= 0.76$ mag. Thus, while the SE knot may not be associated with IC 883, the
blue color of the knot rules it out as a background source. If the knot has
a mass of $\sim 6\times10^6$ M$_\odot$, the estimated age under the
assumption that $A_V = 0$ is 6--7 Myr.  The most likely explanations for the
nature of the knot are that it is an infalling satellite galaxy, or that
it is an off-nuclear star-forming region
in IC 883 similar to Knot S
in the Antennae Galaxy (Whitmore et al. 1999).  Knot S has $\sim 900$ pc
diameter; photometrically, it has a $M_V \sim -13.81$ mag, a $B-I \sim
0.39$ mag and an estimated age of $7\pm1$ Myr.  An additional 
estimation of the
stellar age and of the emission line characteristics of the luminous knot via
optical Keck spectroscopy will be reported by Chien et al. (2011, in
preparation).

\subsection{Fueling Extended Star Formation in IC 883}

Finally, an issue worth further investigation is motivated by the
similarities in measured fluxes of the interferometric and single-dish
measurements of  IC 883.  If 20\% calibration errors for all of the
measurements are assumed, then the majority (or all) of the CO emission in
IC 883 is traced by the interferometric data. I.e., there is little CO
emission beyond the inner 2 kpc of IC 883. This small amount of extended
emission thus must fuel the bulk of the optically-visible star formation,
which accounts for a small fraction of the star formation occurring in IC
883 (i.e., a star formation rate of 0.6 M$_\odot$ yr$^{-1}$ based on the
far-UV emission: Howell et al. 2010).  More sensitive millimeter-wave
observations with the IRAM Plateau de Bure Millimeter Array may reveal the
molecular clouds associated with the optically luminous regions of star
formation observed in the extended regions of IC 883.

\section{Conclusions}

Multi-wavelength GOALS observations of the LIRG IC 883 are presented. These
observations have provided the most well-rounded view to date of the
activity in IC 883. The following conclusions are reached:

{\it (i)} The luminosity function of the star clusters have $\alpha
\sim -2.17\pm0.22$ and $-2.01\pm0.21$ at F435W and F814W, respectively.
These values are within the range of $V$-band $\alpha$
values measured for the well-studied LIRG NGC 34 and the Antennae Galaxy.

{\it (ii)} The colors and absolute magnitudes of the majority of the
star clusters are consistent with instantaneous burst population synthesis
models in which the cluster masses are $10^5$ M$_\odot$ and ages are in the
range of a few$\times10^7 - 10^8$ yrs. Similar models run by making use to
the NICMOS F160W data to derive (F435W - F814W) vs. (F814W - F160W) yield
clusters ages in the range of  $\sim 50-100\times10^6$ years and $> 10^8$
years.  If an $A_V \sim 1.2$ is adopted, which is consistent with
color excesses in LIRGs with measurements of H$\alpha$ and H$\beta$ at $> 3$
kpc and the derived $A_V$ values for clusters in ESO 350-IG038, 
the estimated ages are calculated to be as low as few $\times10^{6-7}$
years.

{\it (iii)} The highest density of optically visible clusters are observed
in an arc $\sim 4-5$ kpc from the optical and 8$\mu$m core. This peak in
the optical  cluster density also corresponds to the peak in the near-UV
emission from IC 883. Both the color of the clusters seen along the line of
sight to the nuclear core, and the underlying galaxy in the core region,
are suggestive of heavy nuclear extinction. This is also supported by the
fact that the mid- and far-infrared, radio, CO, and X-ray emission peak in
the optically red region.

{\it (iv)} The X-ray hardness ratio and the $L_{\rm X(2-10~keV)} / L_{\rm
FIR(40-120\mu m)}$ are both consistent with star formation being the major
contributor to the hard nuclear X-ray emission from IC 883. However, the
large equivalent width of the Si XIII emission line may be due, in part, to
contributions from AGN photoionized gas.

{\it (v)} The CO, near-IR and radio emission have morphologies consistent
with the nuclear starburst being situated in an edge-on disk 2 kpc in
diameter.  An infrared surface brightness $\sim 2\times10^{11}$ L$_\odot$
kpc$^{-2}$ is derived under the assumption that the extent of the infrared
emission is similar to that of the radio emission. The derived infrared
surface brightness is a factor of 10 less than that of the Orion
star-forming region.

{\it (vi)} The high ionization [Ne V] 14.32$\mu$m emission line is also detected.
The emission is further evidence that an AGN is present -- 
it is determined that the [Ne V] luminosity of IC 883 is too high to be due to solely to
supernova remnants.
In addition, the low equivalent width of the 6.2$\mu$m PAH feature $\sim
0.39\mu$m also consistent with an AGN. The ratio of [Ne V] / [Ne II]
12.8$\mu$m is extremely low at $\sim 0.01$, and the ratio of [O IV]
25.9$\mu$m / [Ne II] 12.8$\mu$m $\sim 0.06$. Thus, the AGN appears to be
energetically weak relative to the nuclear starburst population.

{\it (vii)} While the bulk of the X-ray emission is coincident with the
nucleus, a second X-ray source in IC 883 appears to be associated with a
bright, unresolved, off-nuclear component. The component may represent a
large star-forming region, or an infalling satellite galaxy.

\acknowledgements

The authors thank G. Soutchkova, L-H Chien, A. Leroy and
J. Pizagno for useful discussions and
assistance. We also thank the anonymous referee for many suggestions 
which improved the overall quality and clarity of the manuscript.
FM, TV, JP, ASE and DCK were supported by NSF grant AST
02-06262 and by NASA through grants HST-GO10592.01-A and HST-GO11196.01-A
from the SPACE TELESCOPE SCIENCE INSTITUTE, which is operated by the
Association of Universities for Research in Astronomy, Inc., under NASA
contract NAS5-26555. TV acknowledges support from the IPAC Fellowship
Program.  This research has made use of the NASA/IPAC Extragalactic
Database (NED) which is operated by the Jet Propulsion Laboratory,
California Institute of Technology, under contract with the National
Aeronautics and Space Administration. This work is based, in part, on
observations made with the NASA Galaxy Evolution Explorer. GALEX is
operated for NASA by the California Institute of Technology under NASA
contract NAS5-98034. The National Radio Astronomy Observatory is a facility of the 
National Science Foundation operated under cooperative agreement by Associated 
Universities, Inc.

\begin{deluxetable*}{rrrrccccccc}
\pagestyle{empty}
\tabletypesize{\scriptsize}
\tablenum{3}
\tablewidth{0pt}
\tablecaption{IC 883 Star Cluster Photometry}
\tablehead{
\multicolumn{4}{c}{} &
\multicolumn{3}{c}{F435W} &
\multicolumn{2}{c}{F814W} &
\multicolumn{2}{c}{Colors} 
\nl
%\cline{5-7} \cline{8-9} \cline{10-11} \nl
\multicolumn{1}{c}{R.A.} &
\multicolumn{1}{c}{Dec.} &
\multicolumn{1}{c}{$\Delta D^{\rm a}$} &
\multicolumn{1}{c}{$\Delta D^{\rm a}$} &
\multicolumn{1}{c}{$m_{\rm F435W}$} &
\multicolumn{1}{c}{err} &
\multicolumn{1}{c}{$M_{\rm F435W}$} &
\multicolumn{1}{c}{$m_{\rm F814W}$} &
\multicolumn{1}{c}{err} &
\multicolumn{1}{c}{$m_{\rm F435W}-m_{\rm F814W}$} &
\multicolumn{1}{c}{err}
\nl
\multicolumn{1}{c}{($^\prime$ $^{\prime\prime}$)} &
\multicolumn{1}{c}{($^\prime$ $^{\prime\prime}$)} &
\multicolumn{1}{c}{(arcsec)} &
\multicolumn{1}{c}{(pc)} &
\multicolumn{1}{c}{(mag)} &
\multicolumn{1}{c}{(mag)} &
\multicolumn{1}{c}{(mag)} &
\multicolumn{1}{c}{(mag)} &
\multicolumn{1}{c}{(mag)} &
\multicolumn{1}{c}{(mag)} &
\multicolumn{1}{c}{(mag)} 
}
\startdata

          20 35.699 &            8 14.68 &        8.71 &        4.41 &       18.99 &        0.00 &      -16.27 &       18.21 &        0.00 &        0.75 &        0.00 \\
          20 35.094 &            8 22.51 &        3.39 &        1.72 &       19.95 &        0.00 &      -15.31 &       19.17 &        0.00 &        0.75 &        0.00 \\
          20 35.889 &            8 23.87 &        6.69 &        3.39 &       21.66 &        0.00 &      -13.60 &       20.91 &        0.00 &        0.72 &        0.00 \\
          20 35.352 &            8 16.22 &        6.10 &        3.09 &       21.77 &        0.00 &      -13.49 &       21.04 &        0.00 &        0.70 &        0.00 \\
          20 34.792 &            8 27.58 &        8.91 &        4.52 &       22.46 &        0.00 &      -12.80 &       21.32 &        0.00 &        1.11 &        0.00 \\
          20 35.412 &            8 20.57 &        1.82 &        0.92 &       22.72 &        0.03 &      -12.54 &       19.47 &        0.01 &        3.22 &        0.03 \\
          20 35.896 &            8 16.42 &        8.87 &        4.49 &       22.78 &        0.01 &      -12.48 &       21.83 &        0.01 &        0.92 &        0.01 \\
          20 34.986 &            8 16.14 &        7.76 &        3.93 &       22.91 &        0.00 &      -12.35 &       22.24 &        0.01 &        0.64 &        0.01 \\
          20 34.897 &            8 29.77 &        9.55 &        4.84 &       23.05 &        0.01 &      -12.21 &       22.23 &        0.02 &        0.79 &        0.02 \\
          20 35.418 &            8 24.55 &        2.36 &        1.20 &       23.14 &        0.01 &      -12.12 &       21.10 &        0.01 &        2.01 &        0.01 \\
          20 35.545 &            8 13.87 &        8.76 &        4.44 &       23.42 &        0.01 &      -11.84 &       22.49 &        0.01 &        0.90 &        0.01 \\
          20 35.420 &            8 15.56 &        6.78 &        3.44 &       23.50 &        0.01 &      -11.76 &       21.76 &        0.01 &        1.71 &        0.01 \\
          20 34.661 &            8 37.89 &       17.98 &        9.11 &       23.62 &        0.01 &      -11.64 &       22.71 &        0.01 &        0.88 &        0.01 \\
          20 34.744 &            8 29.04 &       10.31 &        5.23 &       23.70 &        0.01 &      -11.56 &       22.12 &        0.01 &        1.55 &        0.01 \\
          20 34.127 &            8 16.79 &       16.31 &        8.27 &       23.72 &        0.01 &      -11.54 &       22.75 &        0.01 &        0.94 &        0.01 \\
          20 35.081 &            8 24.43 &        4.16 &        2.11 &       23.90 &        0.03 &      -11.36 &       21.69 &        0.01 &        2.18 &        0.03 \\
          20 35.481 &            8 22.51 &        1.45 &        0.73 &       23.90 &        0.03 &      -11.36 &       19.87 &        0.00 &        4.00 &        0.03 \\
          20 35.991 &            8 17.49 &        9.16 &        4.65 &       23.92 &        0.02 &      -11.34 &       23.32 &        0.03 &        0.57 &        0.04 \\
          20 35.291 &            8 27.48 &        5.30 &        2.69 &       24.04 &        0.02 &      -11.22 &       23.19 &        0.02 &        0.82 &        0.03 \\
          20 34.954 &            8 14.71 &        9.13 &        4.63 &       24.05 &        0.01 &      -11.21 &       23.27 &        0.02 &        0.75 &        0.02 \\
          20 35.514 &            8 17.20 &        5.44 &        2.76 &       24.05 &        0.02 &      -11.21 &       22.90 &        0.02 &        1.12 &        0.03 \\
          20 34.992 &            8 25.52 &        5.68 &        2.88 &       24.05 &        0.03 &      -11.21 &       21.92 &        0.01 &        2.10 &        0.03 \\
          20 35.988 &            8 20.81 &        7.88 &        3.99 &       24.08 &        0.01 &      -11.18 &       23.33 &        0.02 &        0.72 &        0.02 \\
          20 35.096 &            8 20.93 &        3.61 &        1.83 &       24.12 &        0.02 &      -11.14 &       23.45 &        0.03 &        0.64 &        0.04 \\
          20 35.235 &            8 21.06 &        2.04 &        1.03 &       24.37 &        0.03 &      -10.89 &       23.35 &        0.08 &        0.99 &        0.09 \\
          20 35.803 &            8 13.45 &       10.44 &        5.29 &       24.43 &        0.02 &      -10.83 &       23.64 &        0.02 &        0.76 &        0.03 \\
          20 36.016 &            8 18.08 &        9.14 &        4.63 &       24.65 &        0.04 &      -10.61 &       23.94 &        0.04 &        0.68 &        0.06 \\
          20 34.420 &            8 38.62 &       20.23 &       10.26 &       24.68 &        0.02 &      -10.58 &       24.04 &        0.02 &        0.61 &        0.03 \\
          20 35.757 &            8 16.12 &        7.90 &        4.00 &       24.84 &        0.06 &      -10.42 &       24.17 &        0.07 &        0.64 &        0.09 \\
          20 35.553 &            8 12.04 &       10.55 &        5.35 &       24.93 &        0.03 &      -10.33 &       24.34 &        0.04 &        0.56 &        0.05 \\
          20 35.469 &            8 14.39 &        8.03 &        4.07 &       24.94 &        0.03 &      -10.32 &       22.57 &        0.01 &        2.34 &        0.03 \\
          20 34.691 &            8 18.01 &        9.38 &        4.76 &       24.95 &        0.02 &      -10.31 &       24.25 &        0.03 &        0.67 &        0.04 \\
          20 35.347 &            8 17.70 &        4.62 &        2.34 &       24.98 &        0.06 &      -10.28 &       23.93 &        0.05 &        1.02 &        0.08 \\
          20 35.291 &            8 12.92 &        9.43 &        4.78 &       24.99 &        0.04 &      -10.27 &       24.24 &        0.03 &        0.72 &        0.05 \\
          20 35.829 &            8 11.78 &       12.05 &        6.11 &       25.01 &        0.03 &      -10.25 &       24.31 &        0.04 &        0.67 &        0.05 \\
          20 34.975 &            8 35.00 &       13.67 &        6.93 &       25.05 &        0.04 &      -10.21 &       23.97 &        0.04 &        1.05 &        0.06 \\
          20 35.424 &            8 18.94 &        3.44 &        1.74 &       25.07 &        0.09 &      -10.19 &       24.67 &        0.30 &        0.37 &        0.31 \\
          20 34.906 &            8 25.24 &        6.47 &        3.28 &       25.08 &        0.06 &      -10.18 &       24.47 &        0.07 &        0.58 &        0.09 \\
          20 35.406 &            8 16.37 &        5.97 &        3.03 &       25.09 &        0.05 &      -10.17 &       24.72 &        0.10 &        0.34 &        0.11 \\
          20 35.900 &            8 13.49 &       11.07 &        5.61 &       25.09 &        0.04 &      -10.17 &       24.49 &        0.05 &        0.57 &        0.06 \\
          20 35.116 &            8 21.21 &        3.28 &        1.66 &       25.12 &        0.06 &      -10.14 &       24.40 &        0.08 &        0.69 &        0.10 \\
          20 35.760 &            8 17.51 &        6.87 &        3.48 &       25.15 &        0.09 &      -10.11 &       24.26 &        0.07 &        0.86 &        0.11 \\
          20 35.213 &            8 23.85 &        2.47 &        1.25 &       25.24 &        0.09 &      -10.02 &       22.74 &        0.08 &        2.47 &        0.12 \\
          20 33.409 &            8  4.12 &       30.27 &       15.35 &       25.26 &        0.03 &      -10.00 &       24.89 &        0.05 &        0.34 &        0.06 \\
          20 35.722 &            8 20.81 &        4.67 &        2.37 &       25.28 &        0.06 &       -9.98 &       24.13 &        0.06 &        1.12 &        0.09 \\
          20 35.602 &            8 15.60 &        7.32 &        3.71 &       25.29 &        0.08 &       -9.97 &       24.90 &        0.12 &        0.36 &        0.14 \\
          20 35.197 &            8 17.79 &        4.97 &        2.52 &       25.31 &        0.05 &       -9.95 &       23.46 &        0.02 &        1.82 &        0.05 \\
          20 34.695 &            8 18.95 &        8.97 &        4.55 &       25.37 &        0.04 &       -9.89 &       24.65 &        0.05 &        0.69 &        0.06 \\
          20 34.886 &            8 21.97 &        5.92 &        3.00 &       25.40 &        0.05 &       -9.86 &       24.97 &        0.10 &        0.40 &        0.11 \\
          20 35.287 &            8 11.72 &       10.65 &        5.40 &       25.41 &        0.05 &       -9.85 &       24.65 &        0.05 &        0.73 &        0.07 \\
          20 35.954 &            8 10.00 &       14.37 &        7.29 &       25.46 &        0.05 &       -9.80 &       24.97 &        0.08 &        0.46 &        0.09 \\
          20 34.617 &            8 28.10 &       11.01 &        5.58 &       25.47 &        0.05 &       -9.79 &       24.45 &        0.04 &        0.99 &        0.06 \\
          20 34.786 &            8 24.83 &        7.67 &        3.89 &       25.50 &        0.06 &       -9.76 &       24.70 &        0.06 &        0.77 &        0.09 \\
          20 35.699 &            8 13.03 &       10.21 &        5.17 &       25.50 &        0.05 &       -9.76 &       24.96 &        0.08 &        0.51 &        0.09 \\
          20 34.000 &            8 12.03 &       19.78 &       10.03 &       25.50 &        0.04 &       -9.76 &       24.52 &        0.03 &        0.95 &        0.05 \\
          20 35.177 &            8 18.72 &        4.26 &        2.16 &       25.55 &        0.06 &       -9.71 &       24.69 &        0.08 &        0.83 &        0.10 \\
          20 34.994 &            8 21.96 &        4.64 &        2.35 &       25.57 &        0.07 &       -9.69 &       25.23 &        0.15 &        0.31 &        0.17 \\
          20 36.169 &            8 18.49 &       10.70 &        5.42 &       25.59 &        0.06 &       -9.67 &       24.47 &        0.06 &        1.09 &        0.09 \\
          20 35.005 &            8 30.62 &        9.52 &        4.83 &       25.60 &        0.10 &       -9.66 &       24.98 &        0.16 &        0.59 &        0.19 \\
          20 34.938 &            8 33.94 &       12.88 &        6.53 &       25.60 &        0.08 &       -9.66 &       24.89 &        0.08 &        0.68 &        0.11 \\
          20 35.627 &            8 17.31 &        5.97 &        3.03 &       25.61 &        0.11 &       -9.65 &       24.94 &        0.17 &        0.64 &        0.20 \\
          20 35.669 &            8 27.78 &        6.65 &        3.37 &       25.62 &        0.05 &       -9.64 &       25.01 &        0.08 &        0.58 &        0.09 \\
          20 35.430 &            8 25.27 &        3.09 &        1.57 &       25.64 &        0.08 &       -9.62 &       21.82 &        0.01 &        3.79 &        0.08 \\
          20 35.092 &            8 30.03 &        8.51 &        4.32 &       25.65 &        0.08 &       -9.61 &       25.00 &        0.14 &        0.62 &        0.16 \\
         \enddata
%\tablenotetext{a}{Right ascension in and declination.}
\tablenotetext{a}{Distance of the star cluster from the position of the 8$\mu$m core.}

\end{deluxetable*}
 
\begin{deluxetable*}{rrrrccccccc}
\pagestyle{empty}
\tabletypesize{\scriptsize}
\tablenum{3}
\tablewidth{0pt}
\tablecaption{IC 883 Star Cluster Photometry}
\tablehead{
\multicolumn{4}{c}{} &
\multicolumn{3}{c}{F435W} &
\multicolumn{2}{c}{F814W} &
\multicolumn{2}{c}{Colors} 
\nl
%\cline{5-7} \cline{8-9} \cline{10-11} \nl
\multicolumn{1}{c}{R.A.} &
\multicolumn{1}{c}{Dec.} &
\multicolumn{1}{c}{$\Delta D^{\rm a}$} &
\multicolumn{1}{c}{$\Delta D^{\rm a}$} &
\multicolumn{1}{c}{$m_{\rm F435W}$} &
\multicolumn{1}{c}{err} &
\multicolumn{1}{c}{$M_{\rm F435W}$} &
\multicolumn{1}{c}{$m_{\rm F814W}$} &
\multicolumn{1}{c}{err} &
\multicolumn{1}{c}{$m_{\rm F435W}-m_{\rm F814W}$} &
\multicolumn{1}{c}{err}
\nl
\multicolumn{1}{c}{($^\prime$ $^{\prime\prime}$)} &
\multicolumn{1}{c}{($^\prime$ $^{\prime\prime}$)} &
\multicolumn{1}{c}{(arcsec)} &
\multicolumn{1}{c}{(pc)} &
\multicolumn{1}{c}{(mag)} &
\multicolumn{1}{c}{(mag)} &
\multicolumn{1}{c}{(mag)} &
\multicolumn{1}{c}{(mag)} &
\multicolumn{1}{c}{(mag)} &
\multicolumn{1}{c}{(mag)} &
\multicolumn{1}{c}{(mag)} 
}
\startdata

          20 35.408 &            8 10.38 &       11.96 &        6.07 &       25.66 &        0.05 &       -9.60 &       24.00 &        0.03 &        1.63 &        0.06 \\
          20 32.639 &            7 45.51 &       49.90 &       25.30 &       25.67 &        0.04 &       -9.59 &       25.22 &        0.06 &        0.42 &        0.07 \\
          20 36.138 &            8 23.32 &        9.64 &        4.89 &       25.67 &        0.06 &       -9.59 &       24.91 &        0.08 &        0.73 &        0.10 \\
          20 34.984 &            8 21.37 &        4.84 &        2.46 &       25.69 &        0.08 &       -9.57 &       24.71 &        0.09 &        0.95 &        0.12 \\
          20 34.624 &            8 30.37 &       12.32 &        6.25 &       25.69 &        0.08 &       -9.57 &       24.63 &        0.06 &        1.03 &        0.10 \\
          20 35.023 &            8 17.32 &        6.53 &        3.31 &       25.73 &        0.06 &       -9.53 &       24.96 &        0.06 &        0.74 &        0.09 \\
          20 36.096 &            8 16.27 &       10.93 &        5.54 &       25.74 &        0.08 &       -9.52 &       24.98 &        0.09 &        0.73 &        0.12 \\
          20 34.718 &            8 27.16 &        9.45 &        4.79 &       25.74 &        0.07 &       -9.52 &       24.56 &        0.06 &        1.15 &        0.09 \\
          20 35.858 &            8 17.23 &        7.98 &        4.04 &       25.74 &        0.12 &       -9.52 &       25.05 &        0.14 &        0.66 &        0.18 \\
          20 35.237 &            8 13.98 &        8.48 &        4.30 &       25.74 &        0.06 &       -9.52 &       24.78 &        0.06 &        0.93 &        0.09 \\
          20 34.685 &            8 27.22 &        9.84 &        4.99 &       25.74 &        0.07 &       -9.52 &       25.21 &        0.10 &        0.50 &        0.12 \\
          20 34.878 &            8 15.52 &        9.00 &        4.56 &       25.74 &        0.06 &       -9.52 &       26.05 &        0.20 &       -0.34 &        0.21 \\
          20 35.768 &            8 23.81 &        5.21 &        2.64 &       25.75 &        0.07 &       -9.51 &       24.87 &        0.08 &        0.85 &        0.11 \\
          20 35.893 &            8 14.89 &        9.93 &        5.03 &       25.75 &        0.07 &       -9.51 &       24.70 &        0.06 &        1.02 &        0.09 \\
          20 34.798 &            8 24.47 &        7.41 &        3.76 &       25.78 &        0.08 &       -9.48 &       25.09 &        0.09 &        0.66 &        0.12 \\
          20 34.922 &            8 16.55 &        7.95 &        4.03 &       25.81 &        0.07 &       -9.45 &       25.42 &        0.11 &        0.36 &        0.13 \\
          20 35.585 &            8 29.24 &        7.47 &        3.79 &       25.81 &        0.08 &       -9.45 &       25.07 &        0.09 &        0.71 &        0.12 \\
          20 35.667 &            8 24.98 &        4.66 &        2.36 &       25.81 &        0.08 &       -9.45 &       25.29 &        0.12 &        0.49 &        0.14 \\
          20 34.980 &            8 26.95 &        6.73 &        3.41 &       25.83 &        0.10 &       -9.43 &       24.10 &        0.08 &        1.70 &        0.13 \\
          20 34.836 &            8 28.72 &        9.22 &        4.67 &       25.85 &        0.11 &       -9.41 &       24.64 &        0.09 &        1.18 &        0.14 \\
          20 35.865 &            8 14.09 &       10.33 &        5.23 &       25.86 &        0.07 &       -9.40 &       24.99 &        0.07 &        0.84 &        0.10 \\
          20 35.527 &            8 32.06 &        9.98 &        5.06 &       25.89 &        0.06 &       -9.37 &       25.67 &        0.12 &        0.19 &        0.13 \\
          20 36.094 &            8 17.47 &       10.29 &        5.22 &       25.91 &        0.10 &       -9.35 &       24.57 &        0.07 &        1.31 &        0.12 \\
          20 36.118 &            8 18.48 &       10.11 &        5.13 &       25.92 &        0.12 &       -9.34 &       24.66 &        0.10 &        1.23 &        0.16 \\
          20 35.731 &            8 13.40 &       10.04 &        5.09 &       25.93 &        0.10 &       -9.33 &       24.86 &        0.09 &        1.04 &        0.14 \\
          20 34.015 &            8 37.57 &       22.82 &       11.57 &       25.93 &        0.06 &       -9.33 &       23.87 &        0.02 &        2.03 &        0.06 \\
          20 34.886 &            8 24.56 &        6.40 &        3.25 &       25.93 &        0.10 &       -9.33 &       25.28 &        0.14 &        0.62 &        0.17 \\
          20 34.819 &            8 26.94 &        8.25 &        4.18 &       25.96 &        0.14 &       -9.30 &       25.35 &        0.16 &        0.58 &        0.21 \\
          20 35.175 &            8 30.44 &        8.53 &        4.32 &       25.98 &        0.09 &       -9.28 &       25.17 &        0.10 &        0.78 &        0.14 \\
          20 34.911 &            8 22.44 &        5.66 &        2.87 &       25.98 &        0.10 &       -9.28 &       24.95 &        0.10 &        1.00 &        0.14 \\
          20 35.432 &            8 14.45 &        7.91 &        4.01 &       25.99 &        0.09 &       -9.27 &       25.76 &        0.20 &        0.20 &        0.22 \\
          20 34.707 &            8 21.65 &        8.20 &        4.16 &       26.01 &        0.07 &       -9.25 &       25.00 &        0.07 &        0.98 &        0.10 \\
          20 35.195 &            8 32.03 &       10.02 &        5.08 &       26.05 &        0.09 &       -9.21 &       25.86 &        0.20 &        0.16 &        0.22 \\
          20 33.247 &            7 52.15 &       39.94 &       20.25 &       26.05 &        0.06 &       -9.21 &       24.88 &        0.04 &        1.14 &        0.07 \\
          20 34.653 &            8 34.54 &       15.19 &        7.70 &       26.06 &        0.07 &       -9.20 &       25.57 &        0.12 &        0.46 &        0.14 \\
          20 34.607 &            8 21.20 &        9.48 &        4.80 &       26.06 &        0.08 &       -9.20 &       24.47 &        0.03 &        1.56 &        0.09 \\
          20 35.251 &            8 18.42 &        4.14 &        2.10 &       26.06 &        0.10 &       -9.20 &       24.63 &        0.08 &        1.40 &        0.13 \\
          20 35.106 &            8 17.81 &        5.51 &        2.79 &       26.06 &        0.08 &       -9.20 &       25.56 &        0.13 &        0.47 &        0.15 \\
          20 35.188 &            8 32.58 &       10.58 &        5.36 &       26.07 &        0.10 &       -9.19 &       24.87 &        0.08 &        1.17 &        0.13 \\
          20 34.312 &            8  7.70 &       19.56 &        9.92 &       26.09 &        0.05 &       -9.17 &       25.30 &        0.06 &        0.76 &        0.08 \\
          20 34.762 &            8 21.69 &        7.51 &        3.81 &       26.09 &        0.09 &       -9.17 &       25.76 &        0.14 &        0.30 &        0.17 \\
          20 34.912 &            8 18.05 &        7.04 &        3.57 &       26.11 &        0.09 &       -9.15 &       25.54 &        0.10 &        0.54 &        0.14 \\
          20 34.443 &            8 17.63 &       12.36 &        6.27 &       26.12 &        0.07 &       -9.14 &       25.15 &        0.06 &        0.94 &        0.09 \\
          20 34.549 &            8 24.39 &       10.38 &        5.26 &       26.12 &        0.08 &       -9.14 &       24.80 &        0.05 &        1.29 &        0.09 \\
          20 35.417 &            8 31.92 &        9.68 &        4.91 &       26.13 &        0.10 &       -9.13 &       25.73 &        0.13 &        0.37 &        0.16 \\
          20 35.175 &            8 12.67 &        9.93 &        5.04 &       26.13 &        0.08 &       -9.13 &       25.55 &        0.13 &        0.55 &        0.15 \\
          20 34.826 &            8 34.91 &       14.37 &        7.29 &       26.13 &        0.09 &       -9.13 &       25.24 &        0.10 &        0.86 &        0.14 \\
          20 35.977 &            8 14.14 &       11.18 &        5.67 &       26.14 &        0.11 &       -9.12 &       25.00 &        0.09 &        1.11 &        0.14 \\
          20 33.273 &            7 56.48 &       36.55 &       18.53 &       26.15 &        0.06 &       -9.11 &       25.26 &        0.07 &        0.86 &        0.09 \\
          20 33.480 &            8  4.06 &       29.63 &       15.02 &       26.16 &        0.06 &       -9.10 &       25.44 &        0.08 &        0.69 &        0.10 \\
          20 35.795 &            8 22.10 &        5.33 &        2.70 &       26.16 &        0.11 &       -9.10 &       25.20 &        0.12 &        0.93 &        0.16 \\
          20 36.111 &            8 17.72 &       10.39 &        5.27 &       26.17 &        0.15 &       -9.09 &       24.93 &        0.12 &        1.21 &        0.19 \\
          20 34.989 &            8 35.30 &       13.88 &        7.04 &       26.17 &        0.09 &       -9.09 &       25.21 &        0.12 &        0.93 &        0.15 \\
          20 35.151 &            8 19.28 &        4.02 &        2.04 &       26.18 &        0.11 &       -9.08 &       24.14 &        0.04 &        2.01 &        0.12 \\
          20 35.650 &            8 16.48 &        6.83 &        3.46 &       26.23 &        0.17 &       -9.03 &       25.23 &        0.16 &        0.97 &        0.23 \\
          20 34.752 &            8 21.03 &        7.72 &        3.92 &       26.24 &        0.10 &       -9.02 &       25.80 &        0.15 &        0.41 &        0.18 \\
          20 33.448 &            8  4.12 &       29.89 &       15.16 &       26.25 &        0.07 &       -9.01 &       25.83 &        0.12 &        0.39 &        0.14 \\
          20 35.906 &            8 15.17 &        9.84 &        4.99 &       26.25 &        0.11 &       -9.01 &       25.91 &        0.21 &        0.31 &        0.24 \\
          20 34.243 &            8 15.49 &       15.48 &        7.85 &       26.27 &        0.08 &       -8.99 &       25.66 &        0.10 &        0.58 &        0.13 \\
          20 35.803 &            8  9.71 &       13.78 &        6.98 &       26.27 &        0.10 &       -8.99 &       25.48 &        0.12 &        0.76 &        0.16 \\
          20 35.494 &            8 11.46 &       10.99 &        5.57 &       26.29 &        0.10 &       -8.97 &       25.17 &        0.09 &        1.09 &        0.14 \\
          20 34.782 &            8 28.69 &        9.73 &        4.93 &       26.29 &        0.16 &       -8.97 &       24.77 &        0.10 &        1.49 &        0.19 \\
          20 35.586 &            8 29.51 &        7.73 &        3.92 &       26.35 &        0.12 &       -8.91 &       25.49 &        0.13 &        0.83 &        0.18 \\
          20 36.671 &            8 17.15 &       17.06 &        8.65 &       26.38 &        0.10 &       -8.88 &       25.76 &        0.11 &        0.59 &        0.15 \\
          20 34.588 &            8 29.18 &       11.94 &        6.05 &       26.38 &        0.12 &       -8.88 &       26.10 &        0.22 &        0.25 &        0.25 \\
          20 33.347 &            7 50.35 &       40.55 &       20.56 &       26.41 &        0.09 &       -8.85 &       25.41 &        0.08 &        0.97 &        0.12 \\
          20 35.611 &            8 11.63 &       11.15 &        5.65 &       26.42 &        0.11 &       -8.84 &       25.63 &        0.14 &        0.76 &        0.18 \\
          20 35.145 &            8 34.46 &       12.54 &        6.36 &       26.45 &        0.12 &       -8.81 &       25.39 &        0.12 &        1.03 &        0.17 \\
          20 35.110 &            8 37.33 &       15.43 &        7.83 &       26.47 &        0.12 &       -8.79 &       25.12 &        0.07 &        1.32 &        0.14 \\          
          20 34.626 &            8 23.86 &        9.35 &        4.74 &       26.50 &        0.12 &       -8.76 &       25.89 &        0.16 &        0.58 &        0.20 \\
          20 35.505 &            8 14.93 &        7.59 &        3.85 &       26.51 &        0.15 &       -8.75 &       25.59 &        0.16 &        0.89 &        0.22 \\
          20 34.683 &            8 19.32 &        8.98 &        4.55 &       26.51 &        0.14 &       -8.75 &       25.96 &        0.18 &        0.52 &        0.23 \\
            \enddata
%\tablenotetext{a}{Right ascension in and declination.}
\tablenotetext{a}{Distance of the star cluster from the position of the 8$\mu$m core.}

\end{deluxetable*}

\begin{deluxetable*}{rrrrccccccc}
\pagestyle{empty}
\tabletypesize{\scriptsize}
\tablenum{3}
\tablewidth{0pt}
\tablecaption{IC 883 Star Cluster Photometry}
\tablehead{
\multicolumn{4}{c}{} &
\multicolumn{3}{c}{F435W} &
\multicolumn{2}{c}{F814W} &
\multicolumn{2}{c}{Colors} 
\nl
%\cline{5-7} \cline{8-9} \cline{10-11} \nl
\multicolumn{1}{c}{R.A.} &
\multicolumn{1}{c}{Dec.} &
\multicolumn{1}{c}{$\Delta D^{\rm a}$} &
\multicolumn{1}{c}{$\Delta D^{\rm a}$} &
\multicolumn{1}{c}{$m_{\rm F435W}$} &
\multicolumn{1}{c}{err} &
\multicolumn{1}{c}{$M_{\rm F435W}$} &
\multicolumn{1}{c}{$m_{\rm F814W}$} &
\multicolumn{1}{c}{err} &
\multicolumn{1}{c}{$m_{\rm F435W}-m_{\rm F814W}$} &
\multicolumn{1}{c}{err}
\nl
\multicolumn{1}{c}{($^\prime$ $^{\prime\prime}$)} &
\multicolumn{1}{c}{($^\prime$ $^{\prime\prime}$)} &
\multicolumn{1}{c}{(arcsec)} &
\multicolumn{1}{c}{(pc)} &
\multicolumn{1}{c}{(mag)} &
\multicolumn{1}{c}{(mag)} &
\multicolumn{1}{c}{(mag)} &
\multicolumn{1}{c}{(mag)} &
\multicolumn{1}{c}{(mag)} &
\multicolumn{1}{c}{(mag)} &
\multicolumn{1}{c}{(mag)} 
}
\startdata    
          20 34.756 &            8 36.28 &       16.02 &        8.12 &       26.60 &        0.11 &       -8.66 &       25.61 &        0.11 &        0.96 &        0.16 \\
          20 35.326 &            8 18.40 &        3.93 &        1.99 &       26.62 &        0.23 &       -8.64 &       25.79 &        0.39 &        0.80 &        0.45 \\
          20 33.440 &            8  3.67 &       30.24 &       15.33 &       26.62 &        0.11 &       -8.64 &       25.80 &        0.11 &        0.79 &        0.16 \\
          20 35.306 &            8 12.33 &       10.00 &        5.07 &       26.65 &        0.15 &       -8.61 &       26.00 &        0.20 &        0.62 &        0.25 \\
          20 34.941 &            8 37.09 &       15.79 &        8.01 &       26.67 &        0.13 &       -8.59 &       26.44 &        0.21 &        0.20 &        0.25 \\
          20 33.077 &            7 46.71 &       45.47 &       23.05 &       26.68 &        0.12 &       -8.58 &       26.27 &        0.18 &        0.38 &        0.22 \\
          20 35.968 &            8 25.06 &        7.98 &        4.05 &       26.71 &        0.14 &       -8.55 &       25.66 &        0.13 &        1.02 &        0.19 \\
          20 33.964 &            7 55.00 &       32.36 &       16.41 &       26.71 &        0.11 &       -8.55 &       25.57 &        0.08 &        1.11 &        0.14 \\
          20 36.043 &            8 23.36 &        8.47 &        4.29 &       26.73 &        0.14 &       -8.53 &       25.44 &        0.12 &        1.26 &        0.18 \\
          20 34.281 &            7 59.78 &       26.21 &       13.29 &       26.82 &        0.13 &       -8.44 &       25.51 &        0.08 &        1.28 &        0.15 \\         
          20 33.286 &            7 45.82 &       44.62 &       22.62 &       26.84 &        0.11 &       -8.42 &       25.86 &        0.11 &        0.95 &        0.16 \\
          20 34.191 &            8 38.22 &       21.72 &       11.01 &       26.86 &        0.15 &       -8.40 &       25.54 &        0.09 &        1.29 &        0.17 \\
          20 33.781 &            7 52.17 &       35.93 &       18.22 &       26.87 &        0.12 &       -8.39 &       26.12 &        0.14 &        0.72 &        0.18 \\
          20 35.355 &            8 11.94 &       10.40 &        5.27 &       26.90 &        0.15 &       -8.36 &       25.75 &        0.14 &        1.12 &        0.20 \\
          20 34.992 &            8 18.45 &        6.02 &        3.05 &       26.94 &        0.19 &       -8.32 &       25.34 &        0.10 &        1.57 &        0.22 \\
          20 33.034 &            7 45.66 &       46.60 &       23.63 &       26.94 &        0.13 &       -8.32 &       26.10 &        0.14 &        0.81 &        0.19 \\
          20 36.127 &            8 23.85 &        9.57 &        4.85 &       27.02 &        0.21 &       -8.24 &       24.81 &        0.06 &        2.18 &        0.22 \\
          20 35.138 &            8 37.69 &       15.72 &        7.97 &       27.03 &        0.19 &       -8.23 &       25.43 &        0.09 &        1.57 &        0.21 \\
          20 33.365 &            7 52.69 &       38.59 &       19.56 &       27.15 &        0.16 &       -8.11 &       25.80 &        0.12 &        1.32 &        0.20 \\
          20 36.659 &            8 22.99 &       16.06 &        8.14 &       27.15 &        0.22 &       -8.11 &       25.43 &        0.11 &        1.69 &        0.25 \\
\enddata
%\tablenotetext{a}{Right ascension in and declination.}
\tablenotetext{a}{Distance of the star cluster from the position of the 8$\mu$m core.}

\end{deluxetable*}

\end{document}